\documentclass[12pt]{iopart}

\usepackage{graphicx,bm,url,color}
\usepackage{cite,graphicx,bm,url}
\graphicspath{{./fig/}{./png/}}
\newcommand{\EQ}{\begin{equation}}
\newcommand{\EN}{\end{equation}}
\newcommand{\EQA}{\begin{eqnarray}}
\newcommand{\ENA}{\end{eqnarray}}

\newcommand{\EEq}[1]{Equation~(\ref{#1})}
\newcommand{\Eq}[1]{equation~(\ref{#1})}
\newcommand{\Eqs}[2]{equations~(\ref{#1}) and~(\ref{#2})}
\newcommand{\EEqs}[2]{Equations~(\ref{#1}) and~(\ref{#2})}

\newcommand{\Sec}[1]{Sect.~\ref{#1}}

\newcommand{\Fig}[1]{figure~\ref{#1}}
\newcommand{\FFig}[1]{Figure~\ref{#1}}
\newcommand{\Figs}[2]{figures~\ref{#1} and \ref{#2}}

\newcommand{\bra}[1]{\langle #1\rangle}

\newcommand{\mean}[1]{\overline #1}
\newcommand{\meanrho}{\overline{\rho}}

{}
{}
{}

{}
{}
{}
{}
{}
{}
{}
{}
{}
{}
{}
{}
{}
{}
{}
{}
{}
\newcommand{\meanUU}{\overline{\bm{U}}}

{}

\newcommand{\meanB}{\overline{B}}

\newcommand{\meanU}{\overline{U}}

\newcommand{\meanp}{\overline{p}}

\newcommand{\meanFFF}{\overline{\cal F}}

{}

{}
{}
{}

\newcommand{\meanBB}{{\overline{\bm{B}}}}
\newcommand{\meanJJ}{{\overline{\bm{J}}}}

\newcommand{\bb}{\bm{b}}

\newcommand{\grav}{\bm{g}}

\newcommand{\JJ}{\bm{J}}

\newcommand{\AAA}{\bm{A}}

\newcommand{\UU}{\bm{U}}

\newcommand{\uu}{\bm{u}}

\newcommand{\nab}{{\bm{\nabla}}}

\newcommand{\dd}{{\rm d} {}}

\newcommand{\const}{{\rm const}  {}}

\def\Co{\mbox{\rm Co}}

\def\Pm{\mbox{\rm Pr}_M}
\def\Rm{R_{\rm m}}

\def\Rey{\mbox{\rm Re}}

\def\Co{\mbox{\rm Co}}

\def\Lu{\mbox{\rm Lu}}

\def\cs{c_{\rm s}}

\def\qp{q_{\rm p}}

\def\Peff{{\cal P}_{\rm eff}}
\def\Pmin{{\cal P}_{\rm min}}

\def\pturb{p_{\rm turb}}

\def\qs{q_{\rm s}}

\def\qg{q_{\rm g}}

\def\vA{v_{\rm A}}

\def\kf{k_{\rm f}}

\def\urms{u_{\rm rms}}

\def\etat{\eta_{\rm t}}
\def\etatz{\eta_{\rm t0}}

\def\Beq{B_{\rm eq}}
\def\Beqz{B_{\rm eq0}}
\def\tautd{\tau_{\rm td}}
\def\tauto{\tau_{\rm to}}

\def\half{{\textstyle{1\over2}}}

\def\onethird{{\textstyle{1\over3}}}

\newcommand{\yan}[5]{~ #1~ #5. {\em Astron. Nachr. }{\bf #2}, #3--#4.}
\newcommand{\yana}[5]{~ #1~ #5. {\em Astron. Astrophys. }{\bf #2}, #3--#4.}
\newcommand{\yanaN}[4]{~ #1~ #4. {\em Astron. Astrophys. }{\bf #2}, #3.}

\newcommand{\ypnas}[5]{~ #1~ #5. {\em Proc. Natl. Acad. Sci. }{\bf #2}, #3--#4.}

\newcommand{\yscia}[4]{~ #1~ #4. {\em Sci.\ Adv. }{\bf #2}, #3.}

\newcommand{\ysph}[5]{~ #1~ #5. {\em Solar Phys. }{\bf #2}, #3--#4.}

\newcommand{\yjetp}[5]{~ #1~ #5. {\em Sov. Phys. JETP }{\bf #2}, #3--#4.}

\newcommand{\ysov}[5]{~ #1~ #5. {\em Sov. Astron. }{\bf #2}, #3--#4.}

\newcommand{\ymn}[5]{~ #1~ #5. {\em Monthly Notices Roy. Astron. Soc. }
{\bf #2}, #3--#4.}

\newcommand{\ynat}[5]{~ #1~ #5. {\em Nature }{\bf #2}, #3--#4.}

\newcommand{\yjfm}[5]{~ #1~ #5. {\em J. Fluid Mech. }{\bf #2}, #3--#4.}

\newcommand{\yprs}[5]{~ #1~ #5. {\em Proc. Roy. Soc. Lond. }
{\bf #2}, #3--#4.}
\newcommand{\yps}[4]{~ #1~ #4. {\em Phys. Scr.}{\bf #2}, #3.}

\newcommand{\yapj}[5]{~ #1~ #5. {\em Astrophys. J. }{\bf #2}, #3--#4.}
\newcommand{\yapjN}[4]{~ #1~ #4. {\em Astrophys. J. }{\bf #2}, #3.}

\newcommand{\ypp}[5]{~ #1~ #5. {\em Phys. Plasmas }{\bf #2}, #3--#4.}

\newcommand{\yppN}[4]{~ #1~ #4. {\em Phys. Plasmas }{\bf #2}, #3.}

\newcommand{\ypf}[5]{~ #1~ #5. {\em Phys. Fluids }{\bf #2}, #3--#4.}

\newcommand{\yapjlN}[4]{~ #1~ #4. {\em Astrophys. J. Letters }{\bf #2}, #3.}

\newcommand{\yanf}[5]{~ #1~ #5. {\em Ann. Rev. Fluid Dyn. }{\bf #2}, #3--#4.}

\newcommand{\ypre}[5]{~ #1~ #5. {\em Phys.\ Rev.\ E } {\bf #2}, #3--#4.}
\newcommand{\ylrspN}[4]{~ #1~ #4. {\em Liv.\ Rev.\ Solar Phys.} {\bf #2}, #3.}
\newcommand{\ypreN}[4]{~ #1~ #4. {\em Phys.\ Rev.\ E } {\bf #2}, #3.}
\newcommand{\yjour}[6]{~ #1~ #6. {\em #2} {\bf #3}, #4--#5.}
\newcommand{\yjourN}[5]{~ #1~ #5. {\em #2} {\bf #3}, #4.}

\newcommand{\smn}[2]{ ~#1~ ``#2,'' {\em Monthly Notices Roy. Astron. Soc.} (submitted).}

\begin{document}

\title[
Negative effective magnetic pressure instability
]{
Magnetic concentrations in stratified turbulence:
the negative effective magnetic pressure instability
}

\author{Axel Brandenburg$^{1,2,3,4}$, Igor Rogachevskii $^{5,3}$ and
Nathan Kleeorin $^{5,3}$
}

\address{
$^1$Laboratory for Atmospheric and Space Physics,
    University of Colorado, Boulder, CO 80303, USA\\
$^2$JILA and Department of Astrophysical and Planetary Sciences,
    Box 440, University of Colorado, Boulder, CO 80303, USA\\
$^3$Nordita, KTH Royal Institute of Technology and Stockholm University,
    Roslagstullsbacken 23, SE-10691 Stockholm, Sweden\\
$^4$Department of Astronomy, AlbaNova University Center,
    Stockholm University, SE-10691 Stockholm, Sweden\\
$^5$Department of Mechanical Engineering, Ben-Gurion University of the Negev,
    POB 653, Beer-Sheva 84105, Israel
}
\ead{brandenb@nordita.org, \today, $ $Revision: 1.98 $ $}
\vspace{10pt}


\begin{abstract}
In the presence of strong density stratification, hydromagnetic turbulence
attains qualitatively new properties: the formation of magnetic flux
concentrations.
We review here the theoretical foundations of this mechanism in terms of
what is now called the negative effective magnetic pressure instability.
We also present direct numerical simulations of forced turbulence in
strongly stratified layers and discuss the qualitative and quantitative
similarities with corresponding mean-field simulations.
Finally, the relevance to sunspot formation is discussed.
\end{abstract}

%
%
%
%
%

\section{Introduction}

Magnetohydrodynamic (MHD) turbulence has been studied for a long
time, starting with early work on the energy spectrum
\cite{Iro63,Kra65} in the 1960s.
In many subsequent studies, the effect of gravity was either not
considered or it was thought to being just part of the convective driving
of the turbulence.
The idea that gravity itself could be responsible for causing
qualitatively new phenomena in turbulence hardly occurred.
This has changed dramatically in just the last few years.
An important example is the combined action of gravity with
an imposed vertical magnetic field.
This drives cross helicity \cite{RKB11}, which is an important
invariant in ideal MHD.
Another example, that will form the main focus of this review, is the
spontaneous production of large-scale magnetic flux concentrations
in small-scale MHD turbulence by the negative
effective magnetic pressure instability (NEMPI), which has its roots
in early analytic work \cite{KRR89,KRR90,KMR93,KR94,KMR96,RK07}, and
emerged recently as a pronounced effect in direct numerical simulations
(DNS); see Refs.~\cite{BKKMR11,BKR13,KeBKMR12,KBKMR13,KBKR13}.

In MHD turbulence there is another important effect that leads to
the formation of magnetic structures, namely the dynamo instability.
In that case, gravity is unimportant for structure formation, although
it does often play a role in driving turbulence, for example through
convection.
Thus, the dynamo effect must not be confused with the type of
structure formation where gravity is a crucial ingredient.
Furthermore, convection leads to converging downdrafts that enhance
the magnetic field by compression and tend to expel
it from diverging flow regions \cite{GPW77,TWBP98}.

NEMPI, or some similar process, in conjunction with dynamo theory,
is one of the contenders in explaining the surface activity of the
Sun and other stars.
The other main contender is the rising flux tube scenario by
which strong coherent flux tubes are being built in the tachocline
\cite{Caligari,Fan09,Cha10}, which is the shear layer between the
convection zone and the radiative interior.
However, the helioseismic signatures of such a scenario \cite{BBF10}
have not been detected \cite{Birch16}.
Observations are more consistent with a gradual build-up of an
active region on the timescale of one to two days \cite{SRB16}.
An entirely different kinematic process that can form magnetic
concentrations is flux expulsion, by which magnetic fields are expelled
from regions of rapid motion.
A classical example is a convection cell where magnetic field is
swept away from the diverging upflows of granules into intergranular
lanes and vertices \cite{Cla65,W66}.
Results from relatively weakly stratified numerical simulations
of convection can be explained by this process
\cite{TWBP98,KKWM10,TP13}, but its role in the presence of
strong stratification has not yet been studied.
Numerical simulations with realistic surface physics have successfully
produced active region formation from an unstructured initial magnetic
field \cite{SN12}, but it is still a large leap to modeling actual
sunspots \cite{RS11}.

Meanwhile, several simulations have displayed spontaneous magnetic structure
formation.
Some of them involve turbulent convection
\cite{SN12,TWBP98,KBKMR12,KBKKR16,MS16},
or a stably stratified polytropic atmosphere \cite{LBKR14},
so it remains to be clarified,
whether gravity plays a direct role, or whether the magnetic field
concentrations are mainly the result of converging downdrafts.
Other simulations involve forced turbulence in isothermally stratified domains,
where no thermally driven convection is possible
\cite{BKKMR11,BKR13,KeBKMR12,KBKMR13,KBKR13,MBKR14,JBLKR14,JBMKR16},
and yet one finds the formation of large-scale magnetic structures.
What is remarkable is that these structures extend over the scale of
many turbulent eddies.
This property suggests that they should be amenable to a mean-field
treatment involving averaged, effective equations.

A mean-field approach relevant for describing magnetic
effects on the mean flow was developed nearly three decades ago
\cite{KRR89,KRR90,KMR93,KR94,KMR96,RK07}, but only in recent years,
with the assistance of DNS, has it gained sufficient attention.
In the following, we review the essential properties of NEMPI, discuss
analytical approaches to the understanding of the behavior in the presence
of either horizontal or vertical magnetic fields, and then turn to DNS
whose results can be understood in terms of NEMPI.

\section{The physics of NEMPI}
\label{phys-NEMPI}

\subsection{Effective magnetic pressure}
\label{effect-pressure}

In the following, we discuss the formation of magnetic structures
through a reduction of turbulent pressure by the
large-scale magnetic field.
For large magnetic Reynolds numbers
this suppression of the turbulent pressure can be large enough
so that the effective large-scale magnetic pressure (the sum of non-turbulent
and turbulent contributions to the large-scale magnetic pressure)
becomes negative.
The essence of this effect is as follows.
The momentum equation describing the plasma motions reads
\begin{eqnarray}
{\partial\over\partial t}\rho U_i=-{\partial\over
\partial x_j}\Pi_{ij} + \rho \, g_i,
\label{AA1}
\end{eqnarray}
where ${\bm g}$ is the acceleration due to gravity,
\begin{eqnarray}
\Pi_{ij}=\rho \, U_iU_j+\delta_{ij}\left(p
+\frac{1}{2} {\bm B}^2 \right)-B_iB_j-2\nu \rho \, {\sf S}_{ij},
\label{A1}
\end{eqnarray}
is the momentum stress tensor, $\UU$ and ${\bm B}$ are
the velocity and
magnetic fields, $p$ and $\rho$ are the fluid pressure
and density,
$\delta_{ij}$ is the Kronecker tensor, $\nu$ is the
kinematic viscosity, and
${\sf S}_{ij}=\half(\partial_i U_j+\partial_j U_i)-\onethird\delta_{ij}
\nab\cdot\UU$
is the trace-free rate of strain tensor.
We have adopted units where the vacuum permeability $\mu_0$ is set to unity.

Neglecting correlations between velocity and density fluctuations
for low-Mach number turbulence, the averaged momentum
equation is
\begin{eqnarray}
{\partial\over\partial t} \meanrho \, \meanU_i =
-{\partial\over\partial x_j}\overline{\Pi}_{ij}
+ \meanrho \, g_i,
\label{A2}
\end{eqnarray}
where
$\meanrho$ is the mean fluid density,
$\meanUU$ is the mean fluid velocity,
$\overline\Pi_{ij}=\overline\Pi_{ij}^{\rm m}
+\overline\Pi_{ij}^{\rm f}$ is the mean momentum stress
tensor split into contributions resulting
entirely from the mean field (indicated by superscript m)
and those of the fluctuating field (indicated by superscript f).
In the following, the fluctuations of velocity and magnetic field are
defined as $\uu=\UU-\meanUU$ and $\bb={\bm B}-\meanBB$, respectively,
where $\meanBB$ is the mean magnetic field.
The tensor $\overline\Pi_{ij}^{\rm m}$ has the same form as \Eq{A1},
but all quantities have now attained an overbar, i.e.\
\begin{eqnarray}
\overline\Pi_{ij}^{\rm m}=\meanrho \, \meanU_i\meanU_j
+\delta_{ij}\left(\overline{p}+\frac{1}{2} \overline{{\bm B}}^2\right)
-\meanB_i\meanB_j-2\nu \meanrho \, \overline{\sf S}_{ij},
\label{A3}
\end{eqnarray}
where $\overline{p}$ is the mean fluid pressure.
The contributions $\overline\Pi_{ij}^{\rm f}$,
which describe the effect of turbulence on the large-scale Lorentz force,
are determined by
\begin{eqnarray}
\overline\Pi_{ij}^{\rm f}=\meanrho \, \overline{u_iu_j}
+\half\overline{\bb^2} \delta_{ij}
-\overline{b_ib_j}.
\label{A4}
\end{eqnarray}
The turbulent stress tensor, $\overline\Pi_{ij}^{\rm f}$, together
with the stress tensor describing the mean field contributions,
$\overline\Pi_{ij}^{\rm m}$, comprise the total mean momentum tensor.

Let us first consider isotropic turbulence.
The total (hydrodynamic plus magnetic) turbulent pressure $\pturb$
(i.e., the isotropic part of $\overline\Pi_{ij}^{\rm f}$) is,
in this case, given by \cite{LL75,LL84}
\begin{eqnarray}
\pturb = \frac{2}{3} E_{\rm K} + \frac{1}{3} E_{\rm M} ,
\label{A5}
\end{eqnarray}
where $E_{\rm K} =  \frac{1}{2}\rho \, \overline{{\bm u}^2}$
is the kinetic energy density of the turbulent (small-scale) motions,
$E_{\rm M} = \overline{{\bm b}^2}/2$ is the energy
density of the magnetic fluctuations.
The different coefficients in the ``turbulent'' equation of
state, \Eq{A5}, for the total turbulent pressure are caused by the fact
that the contribution of velocity fluctuations to $\pturb$
is determined by the Reynolds stresses, $\overline{u_i u_j}$, which,
for isotropic turbulence, are
$\overline{u_i u_j} = \frac{1}{3}\overline{{\bm u}^2} \, \delta_{ij}
\equiv \frac{2}{3} (E_{\rm K}/\rho) \, \delta_{ij}$.
On the other hand, the contribution of magnetic fluctuations to $\pturb$
is determined by the Maxwell stresses, $M_{ij} \equiv \frac{1}{2}
\overline{{\bm b}^2} \, \delta_{ij} - \overline{b_i b_j}$, where
$\overline{b_i b_j}= \frac{1}{3}\overline{{\bm b}^2} \, \delta_{ij}$
for isotropic turbulence.
In that case we have $M_{ij}= (\overline{{\bm b}^2}/6) \, \delta_{ij}
\equiv \frac{1}{3} E_{\rm M} \, \delta_{ij}$.

In homogeneous turbulence with a uniform large-scale magnetic field,
the total turbulent energy density $E_{T}=E_{\rm K}+E_{\rm M}$ is
constant, because dissipation is compensated by a continuous supply
of energy \cite{KMR96,RK07}, so
\begin{eqnarray}
E_{\rm K}+E_{\rm M} = \const .
\label{A6}
\end{eqnarray}
This implies that a uniform large-scale magnetic field performs
no work on the turbulence.
It can only redistribute energy
between hydrodynamic and magnetic fluctuations.
\EEq{A6} is a steady-state solution for the budget equation for
the total turbulent energy density for a time-independent energy source
of homogeneous turbulence, $I_T$, with a zero mean velocity
and weakly non-uniform mean magnetic field
\begin{eqnarray}
\frac{\partial E_{T}}{ \partial t} = I_T  +
\etat \, \bm\overline{\bm J}^2
- \frac{E_{T}}{\tau_0},
\label{AA6}
\end{eqnarray}
where ${\bm{\overline{J}}}={\bm \nabla} \times \overline{\bm B}$
is the mean current density, $\tau_0$ is the correlation time
of the turbulent velocity field at the integral scale $\ell_0$ of
turbulent motions, and $\etat$ is the turbulent magnetic diffusion.
The last term, $-E_{T} / \tau_0$, in the right-hand side
of \Eq{AA6} determines the dissipation of the total turbulent energy
for the large fluid and magnetic Reynolds numbers.
\EEqs{A5}{A6} allow us to determine the change
of the total turbulent pressure $\delta\pturb$ in terms of the change of
magnetic energy density $\delta E_{\rm M}$,
\begin{eqnarray}
\delta\pturb = - \frac{1}{3} \,  \delta E_{\rm M} .
\label{A7}
\end{eqnarray}
\EEq{A7} implies that the total turbulent pressure is reduced when
magnetic fluctuations are generated ($\delta E_{\rm M} > 0$).

Let us now consider anisotropic turbulence with a preferred direction
parallel to some unit vector $\hat {\bm e}$.
Specifically, we assume the velocity to be given in the form
${\bm u}={\bm u}_{\perp}+u_{z}\hat{\bm e}$, where
${\bm u}_{z}=u_{z} \hat {\bm e}$ and ${\bm u}_{\perp}={\bm u}-{\bm u}_{z}$
are the velocities parallel and perpendicular to $\hat {\bm e}$.
We characterize the degree of anisotropy by the parameter
$\sigma=\overline{{\bm u}_{\perp}^{2}} / 2\overline{{\bm u}_{z}^{2}}-1$.
Thus, for isotropic three-dimensional turbulence we have $\sigma=0$,
while for strongly anisotropic turbulence $\sigma$ is large.
For two-dimensional turbulence (an extremely anisotropic case)
the parameter $\sigma \to \infty$.
We can then write
\begin{eqnarray}
\overline{u_i u_j} = \frac{1}{3+2\sigma}\overline{{\bm u}^2} \, \left[\delta_{ij}
+ \sigma \left(\delta_{ij} -e_i e_j\right)\right],
\label{A8}\\
\overline{b_i b_j} = \frac{1}{3+2\sigma}\overline{{\bm b}^2} \, \left[\delta_{ij}
+ \sigma \left(\delta_{ij} -e_i e_j\right)\right].
\label{A9}
\end{eqnarray}
Substituting \Eqs{A8}{A9} into \Eq{A4},
the isotropic part of the tensor $\overline\Pi_{ij}^{\rm f}$, which yields
the total turbulent pressure for anisotropic turbulence, becomes
\begin{eqnarray}
\pturb=\frac{2(1+2\sigma)}{3+2\sigma} E_{\rm K}+\frac{1}{3+2\sigma} E_{\rm M}.
\label{A10}
\end{eqnarray}
The change of the total turbulent pressure
$\delta\pturb$ can then be written in terms of the change of
magnetic energy density $\delta E_{\rm M}$ as
\begin{eqnarray}
\delta\pturb = - \frac{1+4\sigma}{3+2\sigma} \,  \delta E_{\rm M} .
\label{A11}
\end{eqnarray}
For strongly anisotropic turbulence with $\sigma \gg 1$ we have
\begin{eqnarray}
\delta\pturb = - 2 \,  \delta E_{\rm M} .
\label{A12}
\end{eqnarray}
Therefore, for strongly anisotropic turbulence
the reduction of the total turbulent pressure by the large-scale
magnetic field is six times larger in comparison with that
for isotropic turbulence [see \Eqs{A7}{A12}].

The turbulent stress tensor, $\overline\Pi_{ij}^{\rm f}$,
is split into parts that are independent of the mean magnetic
field (they determine turbulent viscosity and
background turbulent pressure),
and parts that do depend on the mean magnetic field.
In the presence of a non-zero mean magnetic field
only the difference in the stress tensor,
\begin{eqnarray}
\Delta\overline\Pi_{ij}^{{\rm f}}\equiv
\overline\Pi_{ij}^{{\rm f},\overline{B}}
-\overline\Pi_{ij}^{\rm f,0},
\label{A13}
\end{eqnarray}
depends on the mean magnetic field $\meanBB$,
where $\overline\Pi_{ij}^{\rm f,0}=\overline\Pi_{ij}^{\rm f}(\meanBB=0)$.
To parameterize the tensor $\Delta\overline\Pi_{ij}^{{\rm f}}$
we use symmetry arguments, which allow us to construct
a symmetric tensor with two preferential
directions along the mean magnetic field $\hat {\bm \beta} =\meanBB/|\meanBB|$
and the gravity field $\hat {\bm g} = {\bm g}/g$.
Such a symmetric tensor should be a linear combination of symmetric
tensors $\delta_{ij}$, $\hat \beta_i \hat \beta_j$ and
$\hat g_i \hat g_j$.
These arguments yield \cite{RK07}
\begin{eqnarray}
\Delta\overline\Pi_{ij}^{{\rm f}}= \qs\meanBB^2 \hat\beta_i \hat\beta_j
- \, \Big(\half \qp \, \delta_{ij} + \qg \,
\hat g_i \hat g_j \Big) \meanBB^2,
\label{A14}
\end{eqnarray}
where $\qs$, $\qp$, and $\qg$ are functions of magnetic Reynolds and
Prandtl numbers as well as the modulus of the normalized mean field,
$\beta=\meanB/\Beq$. Here, $\meanB=|\meanBB|$ and
$B_{\rm eq}=(\meanrho \, \overline{\uu^2})^{1/2}$
is the equipartition field strength.
Additional contributions to the tensor $\Delta\overline\Pi_{ij}^{{\rm f}}$
involving, for example, the mean current density are possible,
but will not be considered here.

In summary, the effective mean Lorentz force, which takes into
account the turbulence effects, reads
\begin{eqnarray}
\meanrho \, \meanFFF^{\rm M}_i &\equiv& -\nabla_j
\Big(\half\meanBB^2 \delta_{ij} -\meanB_i\meanB_j
+\Delta\overline\Pi_{ij}^{\rm f}\Big)
\nonumber\\
&=& -\half\nabla_i \Big[(1-\qp) \, \meanBB^2\Big]
+ \hat g_i \, \nabla_z \Big(\qg \,  \meanBB^2\Big)
+ \meanBB \cdot \bm\nabla \Big[(1-\qs)
\,\meanBB\Big].
\label{A15}
\end{eqnarray}
\EEq{A15} has been derived using
the spectral $\tau$ approach \cite{KRR90,KMR96,RK07},
the renormalization approach \cite{KR94}
and the quasi-linear approach \cite{BKKR12,RKS12}.
The sum of non-turbulent and turbulent contributions
to the large-scale magnetic pressure determines
the effective magnetic pressure,
\begin{eqnarray}
p_{\rm eff}=\half\left[1-q_{\rm p}(\beta)\right]\meanBB^2 .
\label{A16}
\end{eqnarray}
When the turbulent contribution, $q_{\rm p}(\beta)$, becomes large enough,
i.e.\ $q_{\rm p}(\beta)>1$, the effective magnetic pressure is negative,
and a large-scale instability, namely NEMPI, can be excited
in strongly stratified turbulence.
This effect will be studied in the next subsections.

\subsection{Growth rate of NEMPI for horizontal field}
\label{intab-1}

To elucidate the mechanism of NEMPI for a horizontal
field, we follow Ref.~\cite{KBKMR13}
and consider an equilibrium with zero mean velocity $\meanUU=\bm{0}$
and a weak imposed constant mean horizontal magnetic field,
$\meanBB_0=(0,B_0,0)$.
We use the linearized MHD equations neglecting for simplicity
the terms which are proportional to the turbulent viscosity
and turbulent magnetic diffusivity caused by the electrical
conductivity of the plasma.
To grasp the essence of NEMPI, we assume that we can
apply the anelastic approximation,
$\nab\cdot\meanUU= -\meanUU\cdot\nab\ln\meanrho$,
\EQ
\nab\cdot\meanUU={\meanU_z\over H_\rho},
\label{B1}
\EN
where $\nab\ln\meanrho=(0,0,-1/H_\rho)$
and $H_\rho$ is the density scale height.
We consider the case $H_\rho=\const$ and $\nabla_y=0$ with
a weak imposed magnetic field in the $y$ direction.
Since the mean magnetic field is independent of $y$,
the mean magnetic tension vanishes in the equation of motion, so
\EQ
{\partial\meanUU(t,x,z)\over\partial t} +(\meanUU \cdot\nab)\meanUU=
-{1\over \meanrho} \nab p_{\rm tot}+\grav,
\label{B2}
\EN
where $p_{\rm tot}=\meanp + p_{\rm eff}$ is the sum of
mean gas pressure $\meanp$ and effective magnetic pressure
$p_{\rm eff}$.
We use the $y$-component of the linearized induction equation
\EQ
{\partial{\tilde B_y}\over \partial t}= - B_0 \, \nab\cdot \tilde{\UU} ,
\label{B3}
\EN
where $\tilde{\UU}$ and $\tilde B_y$ are small perturbations of the mean
velocity and magnetic fields.
To eliminate the gradient of the total pressure in \Eq{B2}, we
take twice the curl of this equation and linearize, and obtain
\EQ
{\partial\over\partial t} \left(\Delta - {1\over H_\rho}\nabla_z\right)
\tilde U_z=2{\vA^2\over H_\rho}\left({\dd \Peff\over\dd\beta^2} \right)_{\beta=\beta_0}
{\nabla_x^2\, \tilde B_y\over B_0} ,
\label{B5}
\EN
where $\Peff(\beta)= p_{\rm eff}/\Beq^2$,
and perturbations of the effective magnetic pressure are
\begin{eqnarray}
\tilde p_{\rm eff} = 2 \, (\meanBB_0 \cdot \tilde {\bm B})
\left({\dd \Peff\over\dd\beta^2} \right)_{\beta=\beta_0}.
\label{CC4}
\end{eqnarray}
We introduce a new variable: $V_z=\sqrt{\meanrho}\, \tilde U_z$,
and use \Eqs{B1}{B3},
\EQ
{\partial\over\partial t} \tilde B_{y}=-B_0{\tilde U_{z}\over H_\rho} ,
\label{B5b}
\EN
which yield
\EQ
{\partial^2\over\partial t^2} \left(\Delta - {1 \over 4 H_\rho^2}\right) V_z(t,x,z)=
- {2\vA^2 \over H_\rho^2} \left({\dd \Peff\over\dd\beta^2} \right)_{\beta=\beta_0} \nabla_x^2 V_z,
\label{B6}
\EN
where $v_{\rm A}(z)=\meanB_0/\sqrt{\meanrho(z)}$
is the mean Alfv\'en speed.
When the characteristic scale of the spatial variation of the
perturbations of the magnetic and velocity fields is much smaller than the
density scale height, $H_\rho$, the growth rate of the instability is
\EQ
\lambda= {\vA \over H_\rho} \,
\left[-2 \left({\dd \Peff\over\dd\beta^2} \right)_{\beta=\beta_0} \right]^{1/2} \, {k_x \over k} \quad\mbox{(horizontal field)}.
\label{B7}
\EN
\EEq{B7} implies that a necessary condition for the large-scale instability is
\EQ
\left({\dd \Peff\over\dd\beta^2} \right)_{\beta=\beta_0} < 0 .
\label{B8}
\EN
In summary, the mechanism of NEMPI with a horizontal weak imposed
magnetic field is as follows.
\EEq{B1} shows that a downward motion $\meanU_z<0$
leads to a compression: $\nab\cdot\meanUU<0$.
This enhances an applied magnetic field locally [see \Eq{B5b}]
and results in a large-scale instability, i.e., NEMPI.
This instability causes the formation of strongly inhomogeneous
magnetic structures.

\subsection{Growth rate of NEMPI for vertical field}
\label{intab-2}

To consider the mechanism of NEMPI for a vertical imposed magnetic field,
$\meanBB_0=(0,0,B_0)$, we follow Ref.~\cite{BGJKR14}.
The MHD equations for small perturbations are
\begin{eqnarray}
&&{\partial \tilde {\bm B}\over \partial t} = (\meanBB_0 \cdot
\nab)\tilde {\bm U} - \meanBB_0 \,\nab\cdot\tilde{\bm U},
\label{C1}\\
&&  {\partial \tilde {\bm U}\over \partial t} =
{1 \over \meanrho} \left[(\meanBB_0 \cdot
\nab)\tilde {\bm B}- \nab \tilde p_{\rm eff}\right],
\label{C3}
 \end{eqnarray}
where
\begin{eqnarray}
\tilde p_{\rm eff} = 2 \, (\meanBB_0 \cdot \tilde {\bm B})
\left({\dd \Peff\over\dd\beta^2} \right)_{\beta=\beta_0}.
\label{C4}
 \end{eqnarray}
We study an axisymmetric problem in cylindrical coordinates,
$(r, \varphi, z)$,
apply the anelastic approximation,
use the magnetic vector potential $A$,
and introduce the stream function $\Psi$, i.e.,
\begin{eqnarray}
\tilde {\bm B} = \nab {\times} \left(A {\rm e}_\varphi \right), \quad
\meanrho \, \tilde {\bm U} = \nab {\times}
\left(\Psi {\rm e}_\varphi \right).
\label{C5}
\end{eqnarray}
Using the radial components of \Eqs{C1}{C3}, we
obtain for $\Phi(t,r,z) = \meanrho^{\,-1} \, \nabla_z \Psi$ the expression
\begin{equation}
{\partial^2 \Phi\over \partial t^2}
=v^2_{\rm A}(z) \left[\nabla_z^2
+2 \left({\dd \Peff\over\dd\beta^2} \right)_{\beta=\beta_0}
\Delta_s \right] \Phi,
\label{C6}
\end{equation}
where $\Delta_s$ is the radial part of the Stokes operator,
\begin{eqnarray*}
\Delta_s = {1 \over r} {\partial \over \partial r}
\left(r {\partial \over \partial r}\right) - {1 \over r^2}.
\end{eqnarray*}
The density profile in an isothermal layer is
$\meanrho=\meanrho_0 \exp (-z/H_\rho)$.
The solution of \Eq{C6} is given by
\begin{eqnarray}
\Phi(t,r,z) = \exp (\lambda t)\, J_1(k_r r) \, \tilde \Phi(z),
\label{C7}
 \end{eqnarray}
where $k_r$ is a suitably defined radial wavenumber and
$J_1(x)$ is the Bessel function of the first kind, which is an
eigenfunction of the radial part of
$\Delta_s$, i.e.,
$\Delta_s J_1(k_r r) = - k_r^2 J_1(k_r r)$.
Substituting \Eq{C7} into \Eq{C6}, we obtain
for $\tilde \Phi(z)$ the equation
\begin{equation}
\nabla_z^2\tilde \Phi(z) - \left[{\lambda^2 \over v^2_{\rm A}(z)}
+2 k_r^2 \left({\dd \Peff\over\dd\beta^2} \right)_{\beta=\beta_0}
\right] \tilde \Phi =0.
\label{C8}
\end{equation}
For $|\nabla_z^2\tilde \Phi(z)|/(k_r^2 \tilde \Phi) \ll 1$,
the growth rate of NEMPI for a vertical imposed field is given by
\begin{eqnarray}
\lambda=v_{\rm A} k_r \left[
-2\left(\dd \Peff \over \dd\beta^2 \right)_{\beta=\beta_0}\right]^{1/2}
\quad\mbox{(vertical field)}.
\label{C10}
\end{eqnarray}
A necessary condition for NEMPI with vertical field is,
again, given by \Eq{B8}.
In summary, the mechanism of NEMPI for a vertical imposed field is as follows.
The downflow removes gas from the upper parts of the turbulent region
so that the pressure decreases, which results in a
return flow that draws with it more vertical field.
This can lead to magnetic field amplification to a strength that exceeds
the equipartition field strength in the upper part of the turbulent
region \cite{BKR13}.

\section{Early simulations toward NEMPI}

\begin{figure}[t!]\begin{center}
\includegraphics[width=\columnwidth]{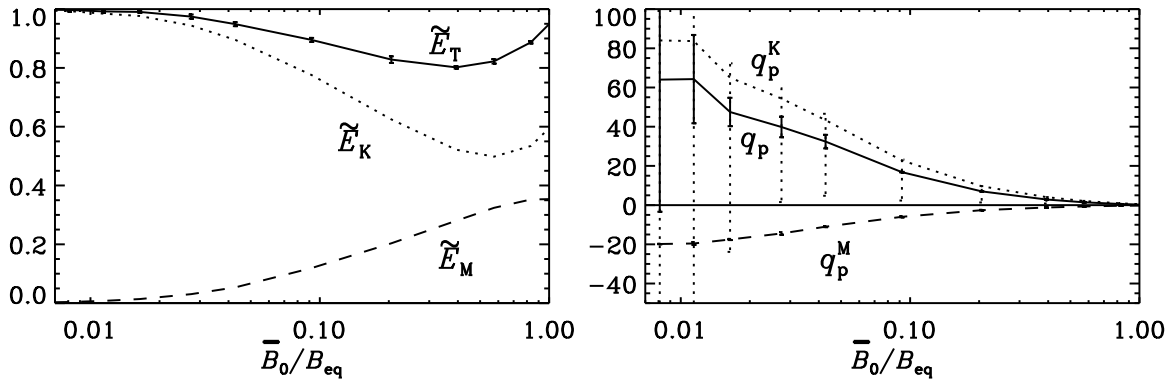}
\end{center}\caption[]{
$\meanB_0$ dependence of the
normalized turbulent energy $\tilde{E}_{\rm T}=E_{\rm T}/E_0$,
where $E_0$ is the value of $E_{\rm T}$ for $\meanB_0=0$,
together with the contributions from kinetic and magnetic energies,
$\tilde{E}_{\rm K}=E_{\rm K}/E_0$ and $\tilde{E}_{\rm M}=E_{\rm M}/E_0$
(left panel), as well as the coefficient $q_{\rm p}$
together with the contributions from velocity,
$q_{\rm p}^{\rm K}$ (dotted line), and magnetic,
$q_{\rm p}^{\rm M}$ (dashed line), fluctuations (right panel),
obtained from DNS for $\Rey=180$ and $\Rm=45$.
Adapted from Ref.~\cite{BKR10}.
}\label{pP64_pq}\end{figure}

Numerical simulations of NEMPI began in 2010 \cite{BKR10}.
An immediate goal of those early simulations was the demonstration
of an instability in a strongly stratified layer using a mean-field
parameterization.
Another goal was the verification of the foundations of NEMPI using DNS.
The main effect that the effective magnetic pressure can be negative,
is demonstrated even in the absence of stratification.
Thus, DNS in triply periodic domains with an imposed magnetic field
have been used.
As the imposed magnetic field is increased,
the turbulence becomes suppressed
(i.e., the turbulent kinetic energy $E_{\rm T}$ decreases), while
the energy of the magnetic fluctuations $E_{\rm M}$ increases;
see the left panel of \Fig{pP64_pq}.

Note that the total turbulent energy is actually not quite constant
as theoretically expected (see \Sec{effect-pressure}),
but it shows a small dip.
This is because for larger fields the turbulent correlation time becomes
weakly dependent on the mean magnetic field.
However, even if the energy were constant, the applied magnetic field
would always have a negative effect on the total (magnetic plus kinetic)
turbulent pressure (see \Sec{effect-pressure}).
The resulting dependence of $\qp$ on $\meanB_0/\Beq$
together with the contributions from velocity fluctuations,
$q_{\rm p}^{\rm K}$, and magnetic fluctuations, $q_{\rm p}^{\rm M}$,
is shown in the right panel of \Fig{pP64_pq}.
The measured DNS parameter $\qs$ has always turned out to be
compatible with zero \cite{BKR10,BKKR12,KBKMR12}.
We shall therefore ignore this term in future considerations.
The value of $\qg$ is usually also found to be negligible. Simulations
of convection suggest that $\qg$ is positive and of the order of 100, but
such a seemingly large values has still only a minor effect in mean-field
simulations (MFS) \cite{KBKMR12}.

A horizontal magnetic field $\bm{B}_0=(0,B_y,0)$ is imposed by writing
the field as $\bm{B}=\bm{B}_0+\nab\times\AAA$ and evolving only the magnetic
vector potential $\AAA$.
The appropriate boundary condition in that case is the perfect conductor
boundary condition, $A_x=A_y=A_{z,z}=0$.
This condition is applied both in the present MFS
as well as in the corresponding DNS discussed below.

\begin{figure}[t!]\begin{center}
\includegraphics[width=.465\columnwidth]{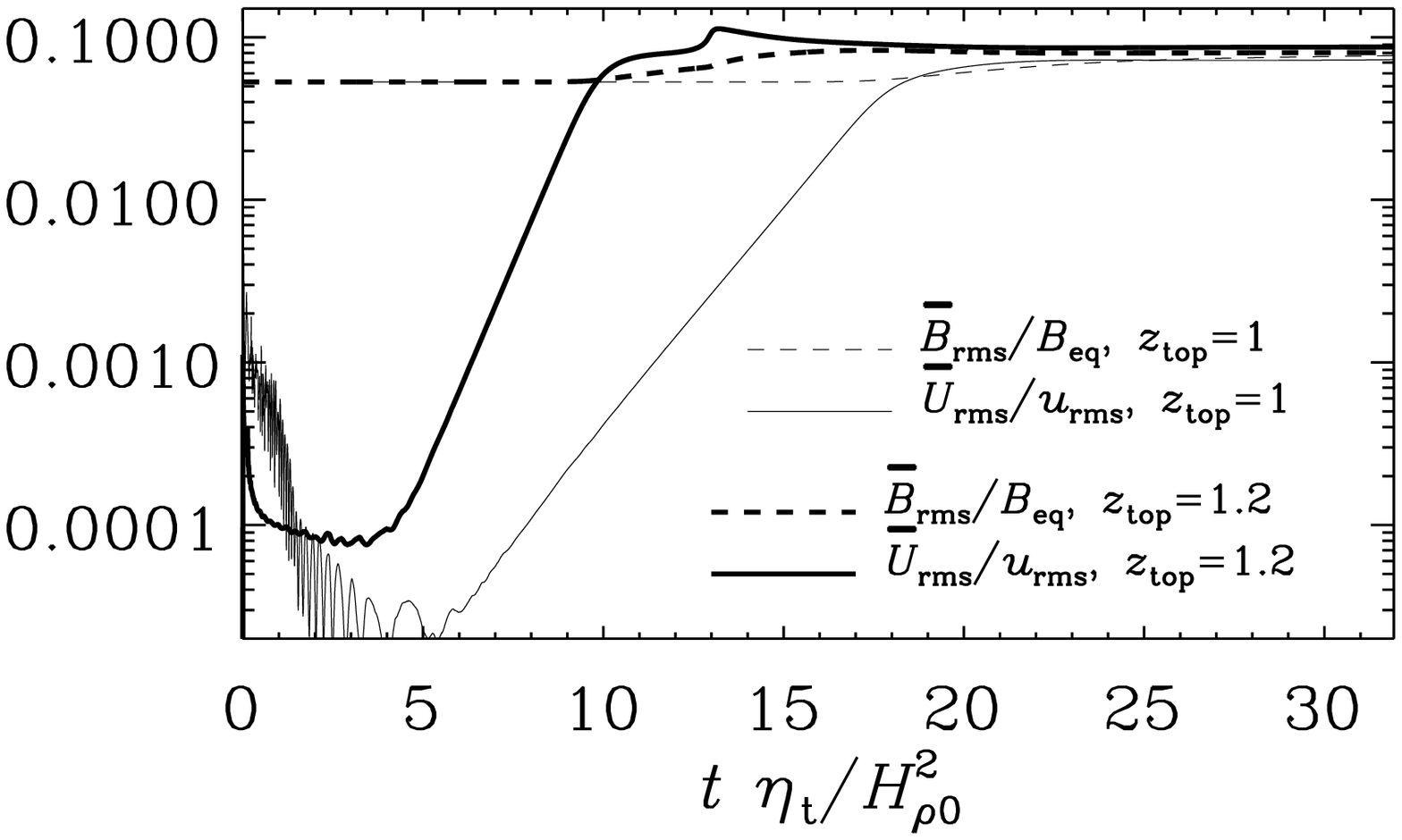}
\includegraphics[width=.525\columnwidth]{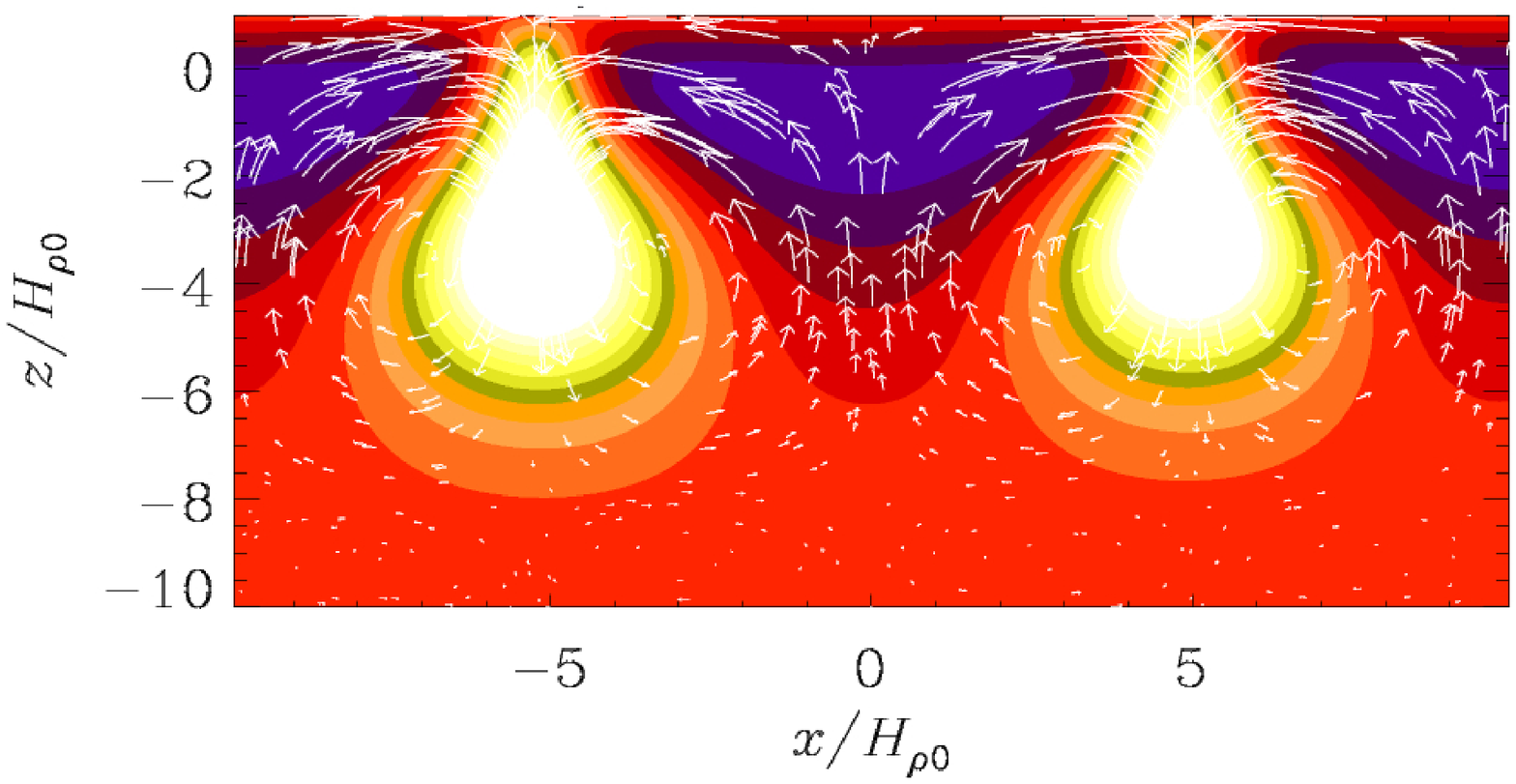}
\end{center}\caption[]{
Left: Growth of the rms value of mean velocity and mean magnetic field
for two runs with different degree of stratification.
Right: Spontaneous production of magnetic flux structures: early
evolution of magnetic field in the $y$ direction (color coded)
together with velocity vectors in the $xz$ plane.
Adapted from Ref.~\cite{BKR10}.
}\label{pbcomp_64x64_strat1_B1_W1a}\end{figure}

Having estimated $\qp$, we can now use this term in a mean-field model
in which the effective (mean-field) magnetic pressure is parameterized
appropriately by replacing
\EQ
\JJ\times\bm{B}\to\meanJJ\times\meanBB+\nab\left(\half\qp\meanBB^2\right).
\EN
If the value of the turbulent magnetic diffusivity is small enough,
the mean-field system shows an instability (i.e., NEMPI) \cite{BKKR12} which
manifests itself mainly in a growth of the mean flow,
because the equilibrium mean flow velocity is zero;
see left panel of \Fig{pbcomp_64x64_strat1_B1_W1a}.
Perturbations to the mean magnetic field also grow exponentially, but
they are harder to detect directly because of the much stronger imposed
magnetic field.

The instability develops concentrations with a certain horizontal
wavelength $\lambda_\perp\approx10\,H_{\rho0}$ or horizontal wavenumber
$k_\perp=2\pi/\lambda_\perp$; see right panel of
\Fig{pbcomp_64x64_strat1_B1_W1a}.
Here, $H_{\rho0}$ is the density scale height at some reference
depth slightly below the surface.
Thus, we have $k_\perp H_{\rho0}\approx0.6$, which is roughly
consistent with subsequent studies \cite{KeBKMR12,KBKMR13,BGJKR14}.

During the nonlinear evolution of NEMPI,
the magnetic structures shown in \Fig{pbymxztn}
take the form of a droplet and look like a
balloon hanging upside down filled with water.
In earlier papers we referred to the associated downflows
as the ``potato sack'' effect \cite{BKR10,BKKMR11}.
We discuss this effect in more detail later in Sect.~\ref{vert}.

\section{Detection of NEMPI in DNS}

\begin{figure*}\begin{center}
\includegraphics[width=.9\textwidth]{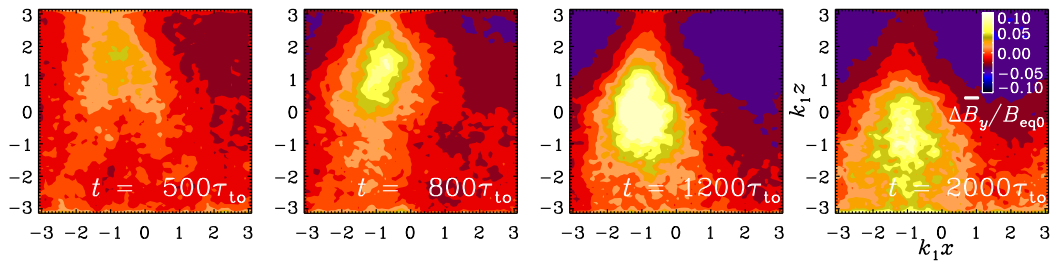}
\end{center}\caption[]{
$(\meanB_y-B_0)/\Beqz$ in the $xz$ plane for magnetic Reynolds number
$\Rm=6$ and imposed field in terms of the equipartition value given by
$B_0/\Beqz=0.05$, showing
a descending potato sack structure.
Time is in turnover times, $\tauto=(\urms\kf)^{-1}$ (lower right).
Adapted from Ref.~\cite{BKKMR11}.
}\label{pbymxztn}\end{figure*}

Let us now turn to the explicit verification of NEMPI using DNS
in strongly stratified forced turbulence.
Early attempts to detect NEMPI failed and it was clear only afterwards
why no NEMPI developed: the scale separation ratio, $\kf/k_1$,
was only about five \cite{KeBKMR12}.
Here, $\kf$ is the forcing wavenumber and $k_1$ is the smallest
wavenumber that fits into the domain.
Simulations with $\kf/k_1=15$ did finally show NEMPI \cite{BKKMR11},
but the effect was still relatively weak and became clearly noticeable only
after having averaged along the direction of the imposed mean magnetic field.
Nevertheless, as seen in \Fig{pbymxztn}, there is a clear sign of the
typical droplet shape associated with subsequent downward motion.

\begin{figure}[t!]\begin{center}
\includegraphics[width=.75\columnwidth]{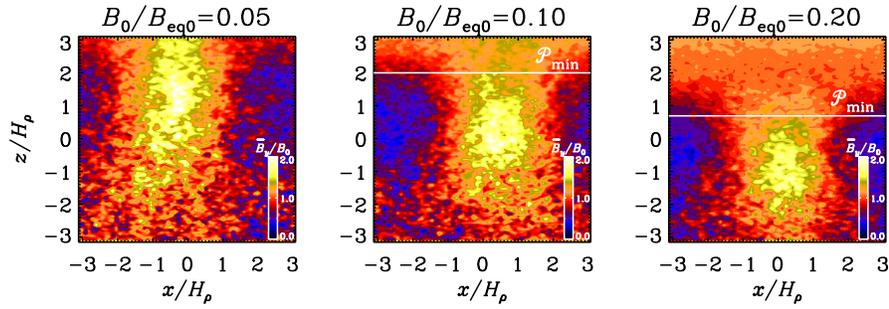}
\end{center}\caption[]{
$\meanB_y/B_0$ from  DNS for three values of the imposed field strength at
the end of the linear growth phase of NEMPI for $\Rm=18$ and $\Pm=0.5$.
The location of the $\Pmin$ line is indicated in panels 2 and 3,
while for panel 1 it lies above the computational domain.
Adapted from Ref.~\cite{KBKMR13}.
}\label{DNS_B}\end{figure}

Another reason for not having noticed NEMPI in earlier DNS studies could be
related to the fact that the field strength was not in the right range.
The  understanding of this aspect came as a benefit of having used
idealizing condition such as an isothermal equation of state.
In that case the scale height is constant and independent of $z$, so
the system is similar at all depths except that the density changes.
For a given imposed field strength $B_0$, there will always be one
particular height where $B_0/\Beq$ takes the preferred value for NEMPI
to develop ($B_0/\Beq$ is in the range $0.03$--$0.2$; see table~1
of Ref.~\cite{LBKR14}).
If the field strength is increased, NEMPI develops simply at a
larger depth \cite{KBKR12}.
This is shown in \Fig{DNS_B}.

\begin{figure}[t!]\begin{center}
\includegraphics[width=.85\columnwidth]{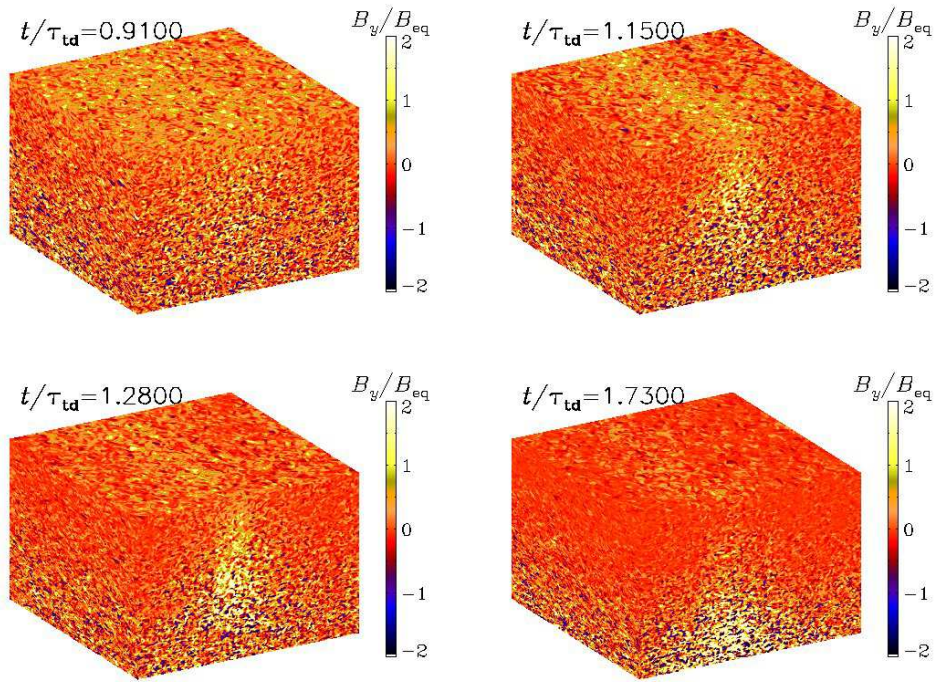}
\end{center}\caption[]{
Visualizations of $B_y$ on the periphery of the domain for different times.
Adapted from Ref.~\cite{KeBKMR12}.
}\label{Brhodos12}\end{figure}

In \Fig{Brhodos12} we show the magnetic field for a run with $\kf/k_1=30$,
so NEMPI is now stronger than before and the flux concentrations can clearly
be seen in snapshots even without averaging.
Here, time is given in turbulent-diffusive times,
$\tautd=(\etatz k_1^2)^{-1}$, where $\etatz=\urms/3\kf$ is the
estimated turbulent magnetic diffusivity.
Again, the field concentration develops first
near the top of the domain and then sinks downward.
Clearly, this is just opposite to the usual magnetic buoyancy instability
\cite{Par67,HP88}, where magnetic fields rise toward the surface.
This underlines the physical reality of a negative effective magnetic
pressure.

\begin{figure}\begin{center}
\includegraphics[width=\columnwidth]{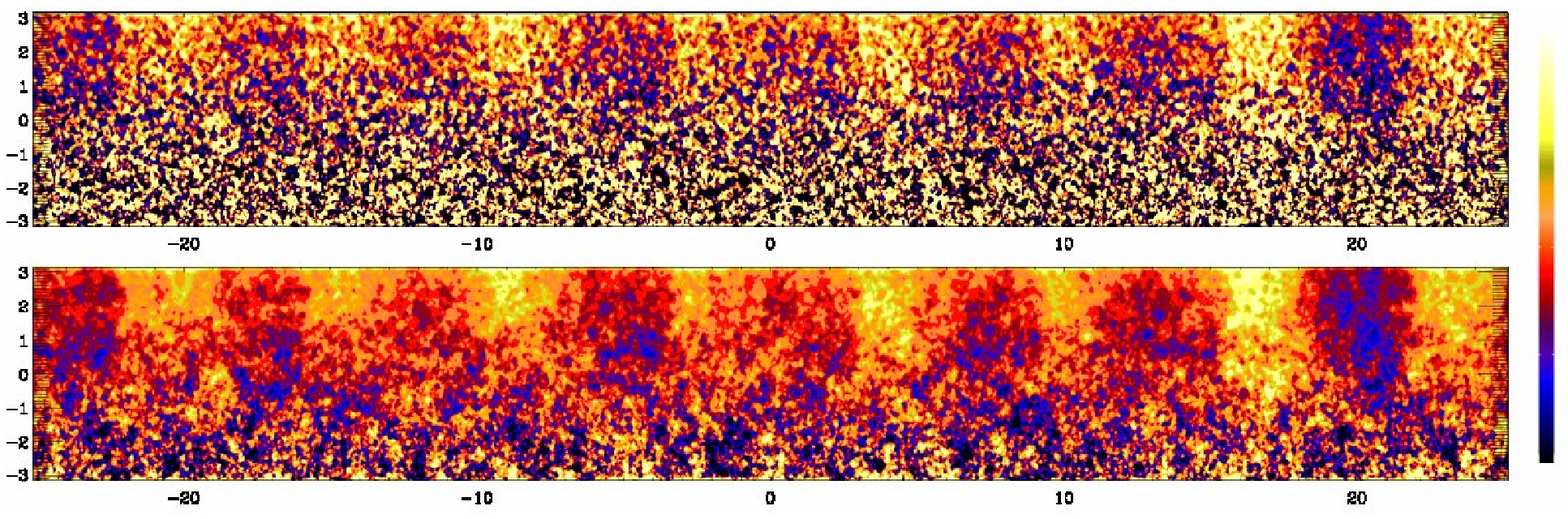}
\end{center}\caption[]{
Visualization of $\meanB_y(x,z)$ for an elongated box
with $\Rm=36$ at a time during the statistically steady state.
The top panel shows the $y$ average $\meanB_y/\Beq$
at one time while the lower panel shows an additional time average
$\bra{\meanB_y}_{t}/\Beq$ covering about 80 turnover times.
The dimensions in the horizontal and vertical directions are $H_\rho$
so the extent is $16\pi H_\rho\times2\pi H_\rho$.
Adapted from Ref.~\cite{KeBKMR12}.
}\label{pbymxz10}\end{figure}

Making the domain wider results in the replication of statistically similar
flux concentrations.
In \Fig{pbymxz10} we see that the wavelength is approximately $8\,H_\rho$,
so $k_\perp H_\rho=2\pi/8\approx0.8$.
Similar behavior is also found in MFS, both in two and three dimensions
\cite{BGJKR14}.

In Ref.~\cite{KBKMR13}, an expression for the growth rate of NEMPI was
derived using the anelastic approximation under the assumption
$H_\rho=\const$ (see for details Sect.~2).
Remarkable agreement with the numerical calculations has been
reported in Ref.~\cite{KBKMR13}, where several examples were shown
that demonstrated quantitative agreement between DNS and MFS; see also
Ref.~\cite{BGJKR14}.

\section{NEMPI with vertical field}
\label{vert}

\begin{figure*}\begin{center}
\includegraphics[width=\textwidth]{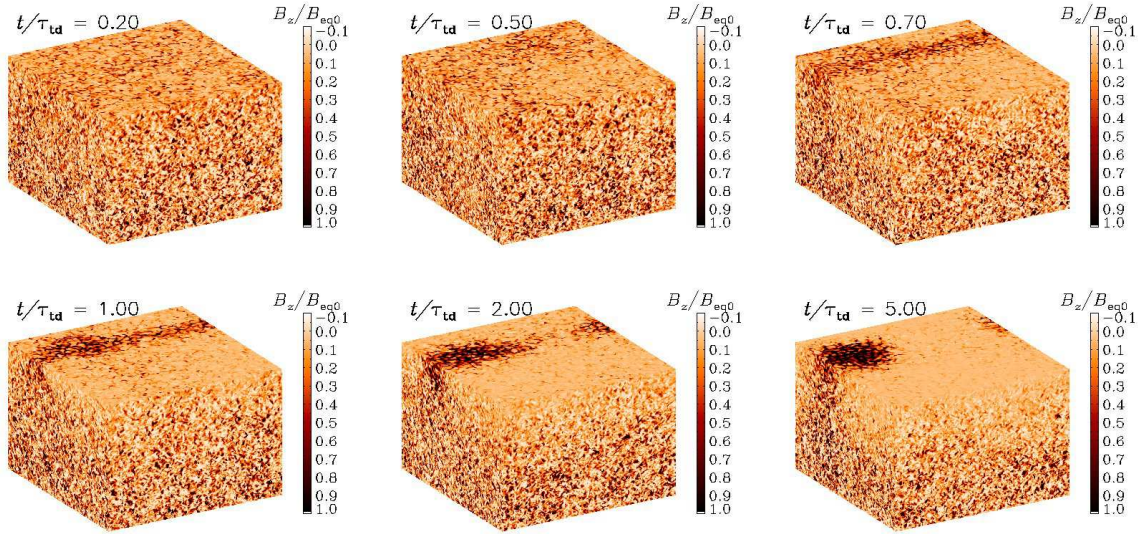}
\end{center}\caption[]{
Evolution from a uniform initial state toward a circular spot
for $B_{z0}/\Beqz=0.02$.
Here, $B_z/\Beqz$ is shown on the periphery of the domain.
Dark shades correspond to strong vertical fields.
Time is in units of $\tautd$.
An animation is available on {\tt http://youtu.be/Um\_7Hs\_1RzA}.
Adapted from Ref.~\cite{BKR13}.
}\label{AB}\end{figure*}

\begin{figure*}\begin{center}
\includegraphics[width=\textwidth]{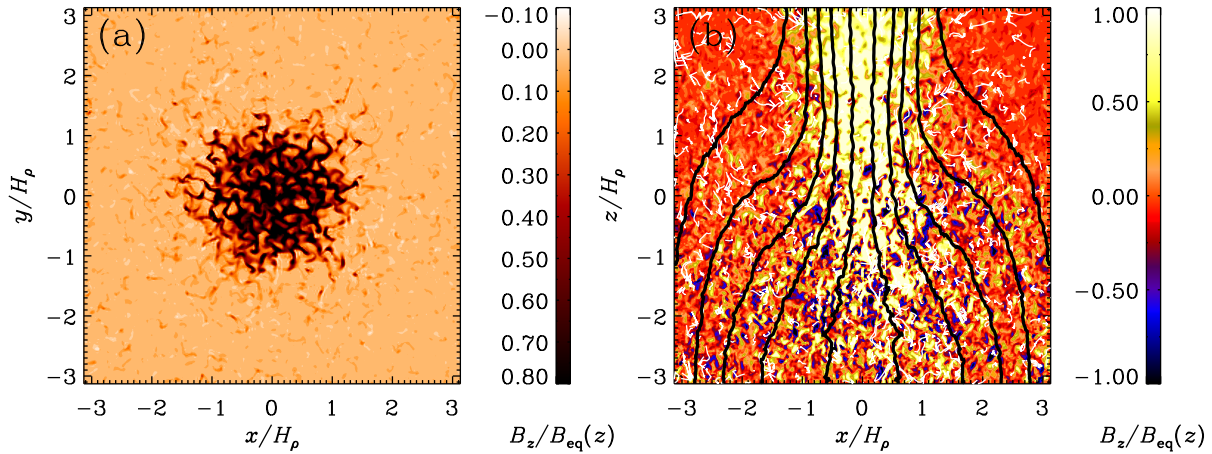}
\end{center}\caption[]{
Cuts of $B_z/\Beq(z)$ in the $xy$ plane at the top boundary
($z/H_\rho=\pi$) and the $xz$ plane through the middle of the spot at $y=0$.
In the $xz$ cut, we also show magnetic field lines and flow vectors
obtained by numerically averaging in azimuth around the spot axis.
Adapted from Ref.~\cite{BKR13}.
}\label{pslices_V256k30VF_Bz002}\end{figure*}

In the presence of a vertically imposed magnetic field,
$\bm{B}_0=(0,0,B_0)$, the appropriate boundary condition is the so-called
vertical field boundary condition, $A_{x,z}=A_{y,z}=A_z=0$.
It turns out that with a vertical field, the effect of
NEMPI is much stronger; see \Fig{AB} for a visualization of $B_z$ on
the periphery of the computational domain at different times.
In this case, $\meanB_z/\Beq$ can reach and even exceed unity.
One reason for this is that the downflows associated with NEMPI lead to a
converging return flow in the upper layers, which pinches the field further.
Cross sections of the resulting magnetic field look quite different from
standard visualizations of buoyant flux tubes that rise and pierce
the surface.
Here, the magnetic field seems to diffuse out as one goes deeper down, see
\Fig{pslices_V256k30VF_Bz002}, where the field lines have been computed
after taking an axisymmetric average of the magnetic field obtained from
the DNS.

\begin{figure}\begin{center}
\includegraphics[width=.352\columnwidth]{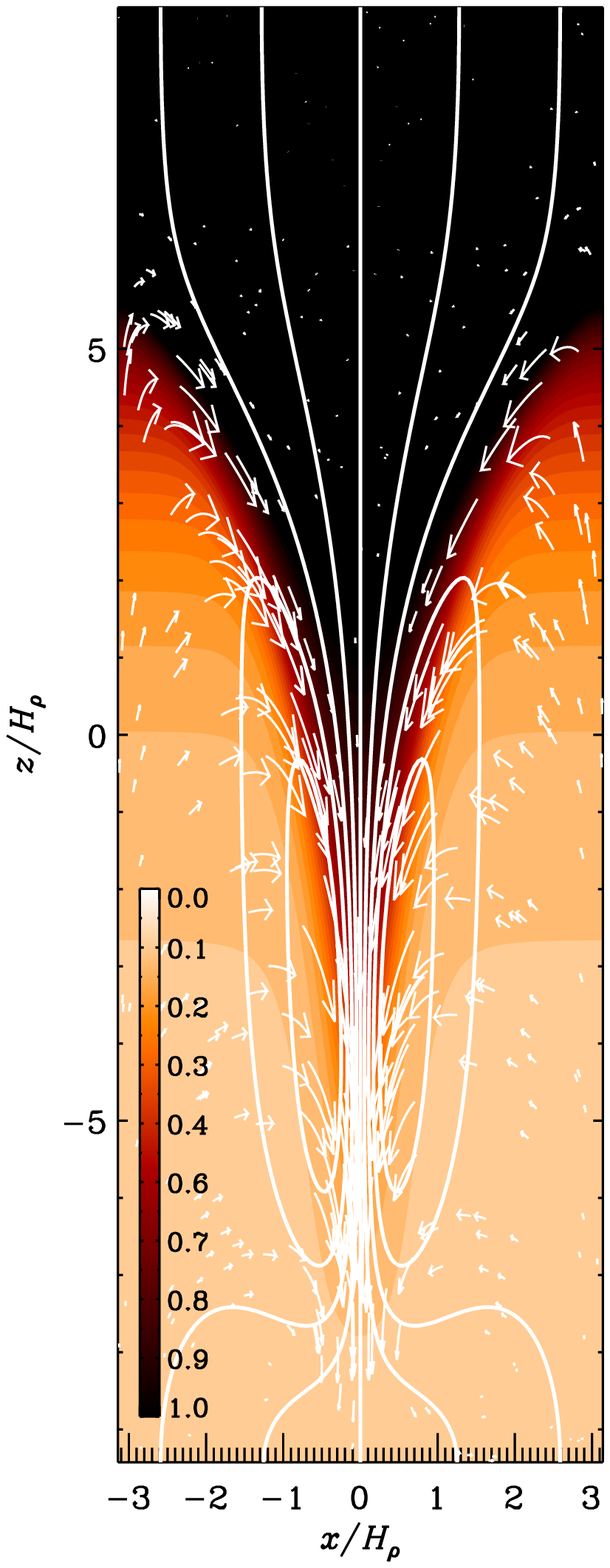}
\includegraphics[width=.536\columnwidth]{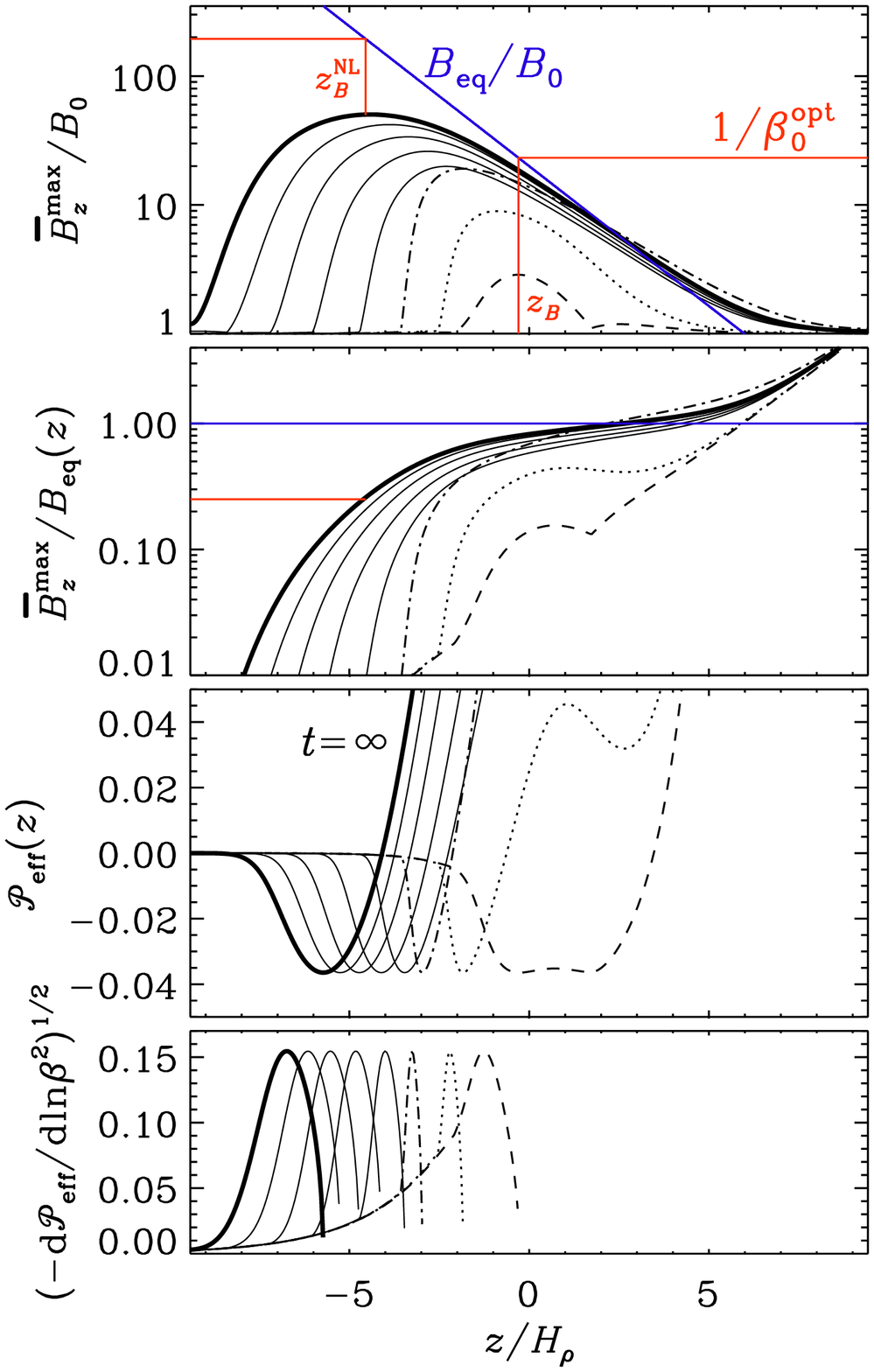}
\end{center}\caption[]{
Left: $\meanB_z/\Beq$ together with field lines and flow vectors from
an axisymmetric MFS with $B_0/\Beqz=0.05$.
The flow speed varies from $-0.27\urms$ (downward) to $0.08\urms$ (upward).
Right: Time evolution of normalized vertical magnetic field profiles,
(a) $\meanB_z^{\rm max}/B_0$ together with
$\Beq(z)/B_0$ (shown by blue line), (b) $\meanB_z^{\rm max}/\Beq(z)$, as well as
(c) $\Peff(z)$ and (d) $(-\dd\Peff/\dd\ln\beta^2)^{1/2}$,
from a MFS with $B_0/\Beqz=0.05$
at $t/\tautd=2.9$ (dashed), 3 (dotted), 3.1 (dash-dotted),
3.3, 3.7, 4.2., 5, and 50 (thick solid line).
The blue solid lines indicate $\Beq(z)$, normalized by (a) $B_0$ and
(b) by itself (corresponding thus to unity).
The red lines indicate the locations $z_B$ and $z_B^{\rm NL}$
during linear and nonlinear (NL) phases of the evolution,
as well as relevant intersections with normalized values of
$\meanB_z^{\rm max}$ and $\Beq$.
Adapted from Ref.~\cite{BGJKR14}.
}\label{pBmax_axi}
\end{figure}

The resulting flux concentrations from NEMPI with a vertical
imposed magnetic field are much stronger compared to the case
with a horizontal field because of the apparent absence of the
aforementioned potato sack effect for vertical field.
This effect has been observed in turbulence with a horizontal
imposed magnetic field; see \Figs{pbymxztn}{DNS_B}.
The potato sack effect is a direct consequence of the negative effective
magnetic pressure, making such magnetic structures heavier than their
surroundings \cite{BKR10,BKKMR11,KBKMR12}.
The potato sack effect removes horizontal magnetic field
structures from regions in which NEMPI is excited and pushes them downward.
For vertical magnetic field, the heavier fluid moves
downward along the field without affecting the flux tube, so
that NEMPI is not stabilized prematurely by the potato sack effect.

In a mean-field framework, it is quite straightforward to produce an
axisymmetric model of a magnetic spot \cite{BGJKR14}.
However, it is important to make sure that the outer radius of the
domain is chosen in a suitable manner.
If it is too big, downflows will develop on the rim of the cylinder;
see figure~8 of Ref.~\cite{BGJKR14}.
Furthermore, for an isothermal gas it is straightforward to extend
the domain arbitrarily in the vertical direction upward and downward.
\FFig{pBmax_axi} shows the resulting flux concentration in a domain
tall enough so that the field becomes uniform both far above and
far below the flux concentration.

\section{Bipolar regions and dynamo-generated magnetic fields}

\begin{figure}[t!]
\begin{center}
\includegraphics[width=0.49\columnwidth]{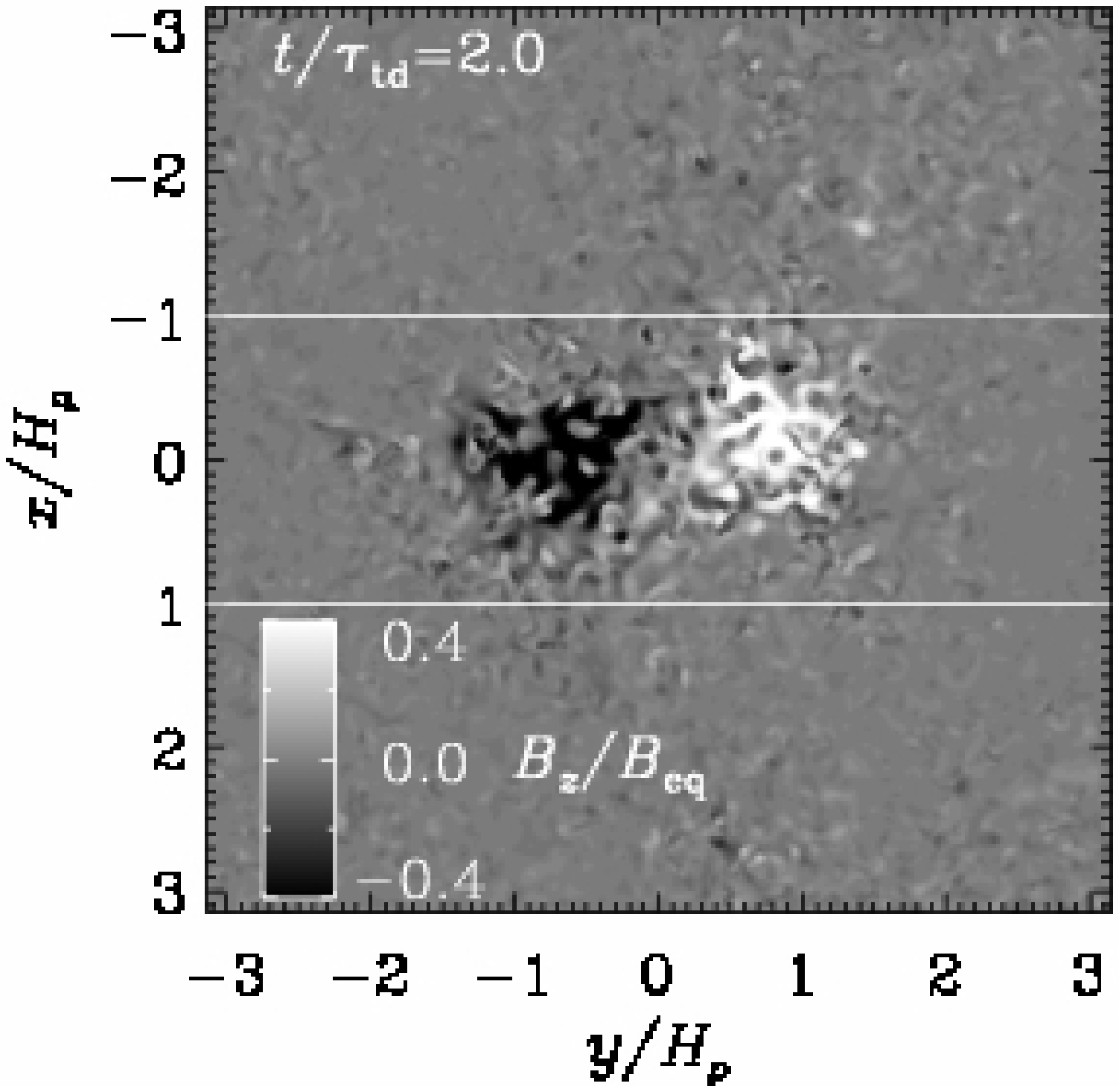}
\includegraphics[width=0.49\columnwidth]{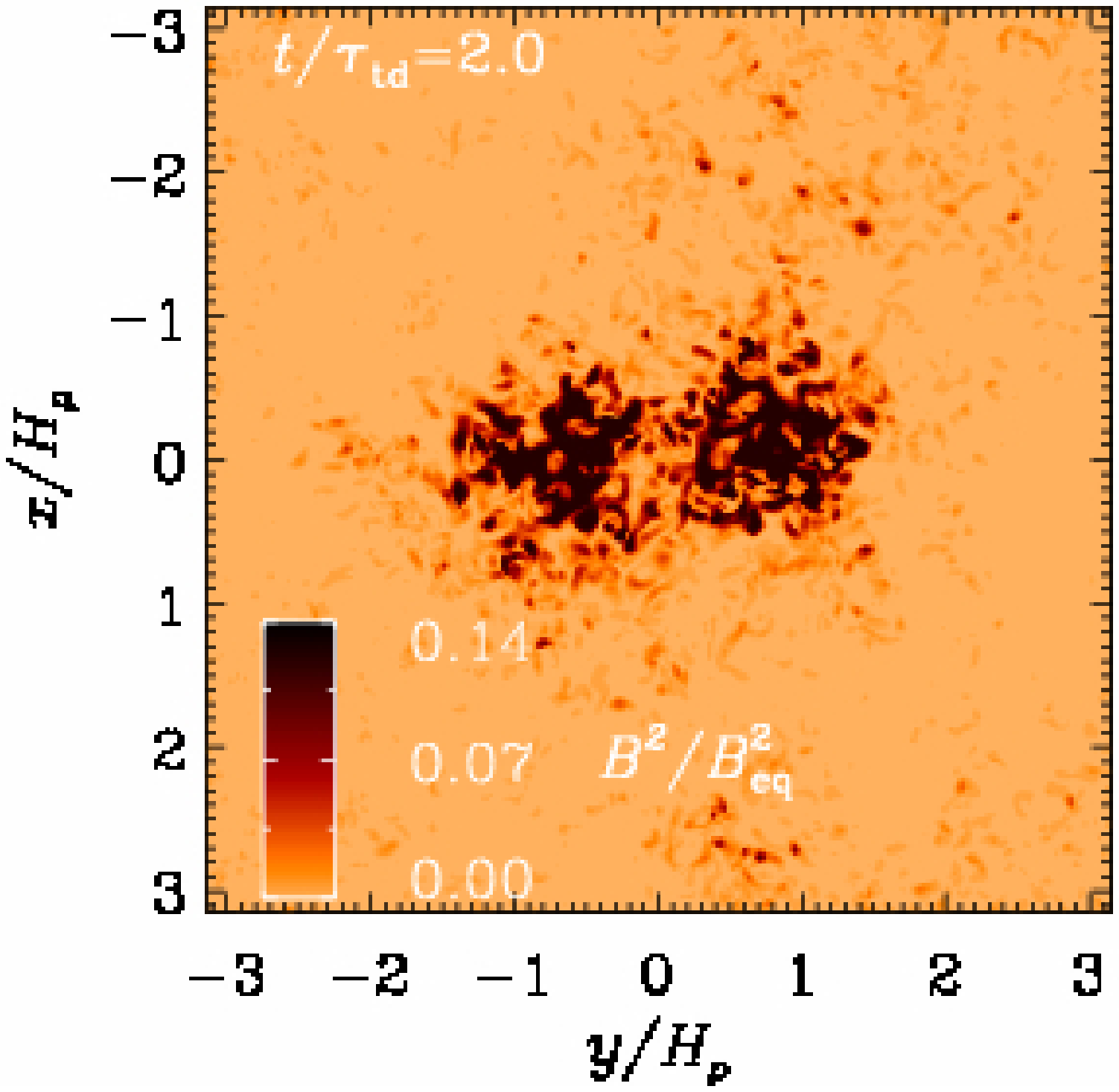}
\end{center}\caption[]{
Left panel: normalized vertical magnetic field $B_z/\Beq$ of the bipolar
region at the surface ($z=0$) of the simulation domain.
Right panel: normalized magnetic energy $\meanBB^2/\Beq^2$ of the
two regions relative to the rest of the surface.
Note that we clip both color tables to increase the contrast of
the structure. The field strength reaches around $B_z/\Beq=1.4$.
Adapted from Ref.~\cite{WLBKR13}.
}\label{b2_xy}
\end{figure}

In reality, there will never be a purely vertical nor a purely horizontal
imposed field, but a dynamo-generated one which always has some natural
horizontal variability.
This would allow the field to penetrate the surface.
Thus, when allowing for a layer above which there is no
turbulence, one can see the development of bipolar regions
\cite{WLBKR13,WLBKR16}.
This layer is meant to represent the effects of a free surface
or a photosphere.
An example of this is shown in \Fig{b2_xy}.
Here, a Cartesian domain of isothermally stratified
gas was divided into two layers.
In the lower layer, turbulence was forced with transverse nonhelical
random waves, whereas in the upper layer no flow was driven.
A weak uniform magnetic field was imposed in the entire domain at all times.
Formation of bipolar magnetic structures
was found over a large range of parameters.
The magnetic structures became more intense for higher stratification until
a density contrast of around 100 across the turbulent layer was reached.
The magnetic field in bipolar regions was found to increase with
higher imposed field strength until the field became comparable to the
equipartition field strength of the turbulence.
A weak imposed horizontal field component turned out to be necessary
for generating bipolar structures.
In the case of bipolar region formation, an exponential growth of the
large-scale magnetic field was found, which is a clear indication of a
hydromagnetic instability \cite{WLBKR16}.
Additionally, the flux concentrations were correlated with strong large-scale
downward-oriented converging flows.
These findings strongly suggest that NEMPI is indeed responsible for
magnetic flux concentrations found in this system.

\begin{figure*}[t!]\begin{center}
\includegraphics[width=\textwidth]{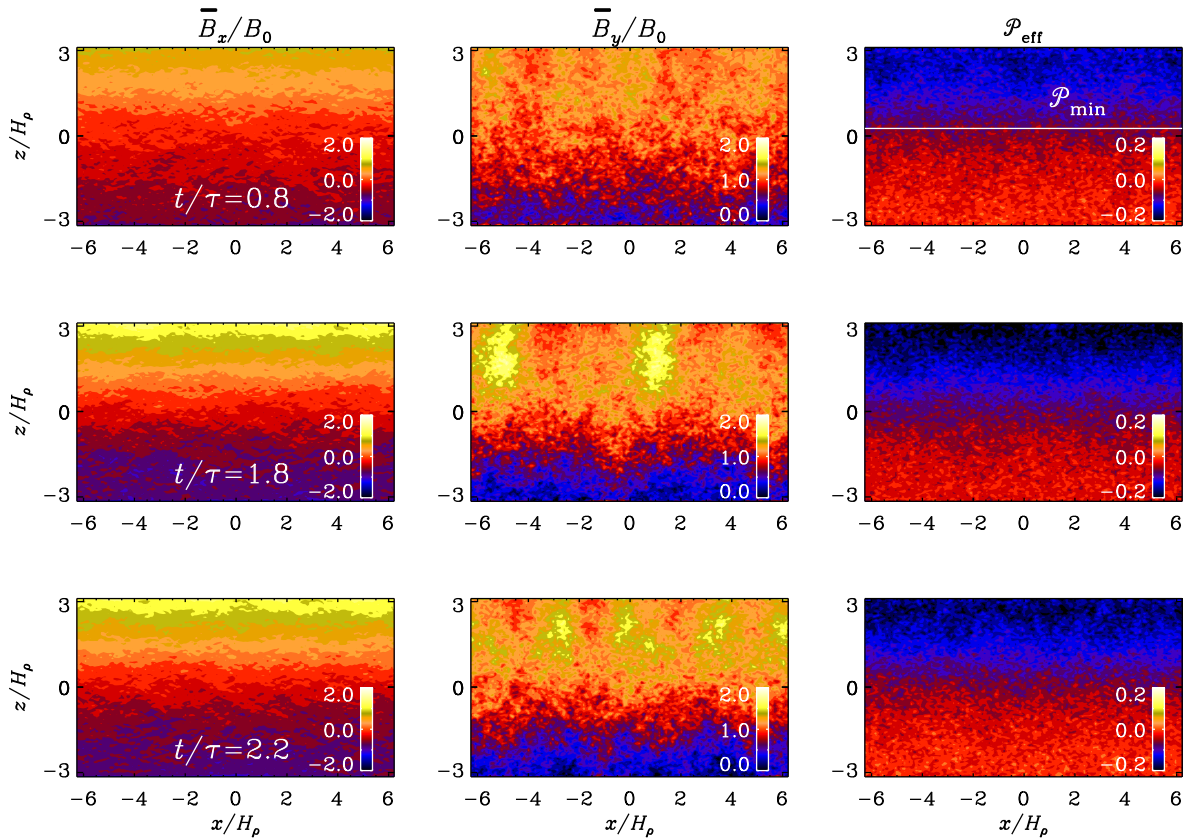}
\end{center}\caption[]{
Visualization of $\meanB_x/B_0$ and $\meanB_y/B_0$ together with
effective magnetic pressure for different times.
Here the angular velocity is $\Omega=0.15$, corresponding to
a Coriolis number $\Co=0.09$.
Adapted from Ref.~\cite{JBLKR14}.
}\label{DNS_P_Om15_b005_th0b_2pi}\end{figure*}

Many subsequent studies of NEMPI have been undertaken in the meantime.
Of particular interest is the case where the main field is not an
imposed  one, but the result of a dynamo.
This situation was first studied in MFS in spherical shells \cite{Jab13},
and later also in Cartesian domains \cite{JBLKR14}.
An example is shown in \Fig{DNS_P_Om15_b005_th0b_2pi}.
Here, the large-scale dynamo is the result of the combined action of
stratification and rotation giving rise to kinetic helicity of the turbulence.
On the other hand, if rotation is too strong, NEMPI will be suppressed
\cite{Los12,Los13}.

The suppression of NEMPI by rotation came as a surprise, especially
because the critical values of the maximum permissible rotation speed
were found to be rather low.
In dynamo theory the importance of rotation on the flow is usually measured
in terms of the Coriolis number, $\Co=2\Omega\tautd$, but those values
are only around 0.1 when NEMPI begins to be suppressed; see figure~2 of \cite{Los13}.
They argued that the reason for this is the fact that the growth time for
NEMPI is longer than the turnover time of the turbulence $\tautd$.
If one normalizes instead with the typical growth rate of NEMPI,
$\lambda_{0*}$ the critical values of $2\Omega/\lambda_{0*}$ are found
to be slightly above unity.

\begin{figure*}\begin{center}
\includegraphics[width=.32\columnwidth]{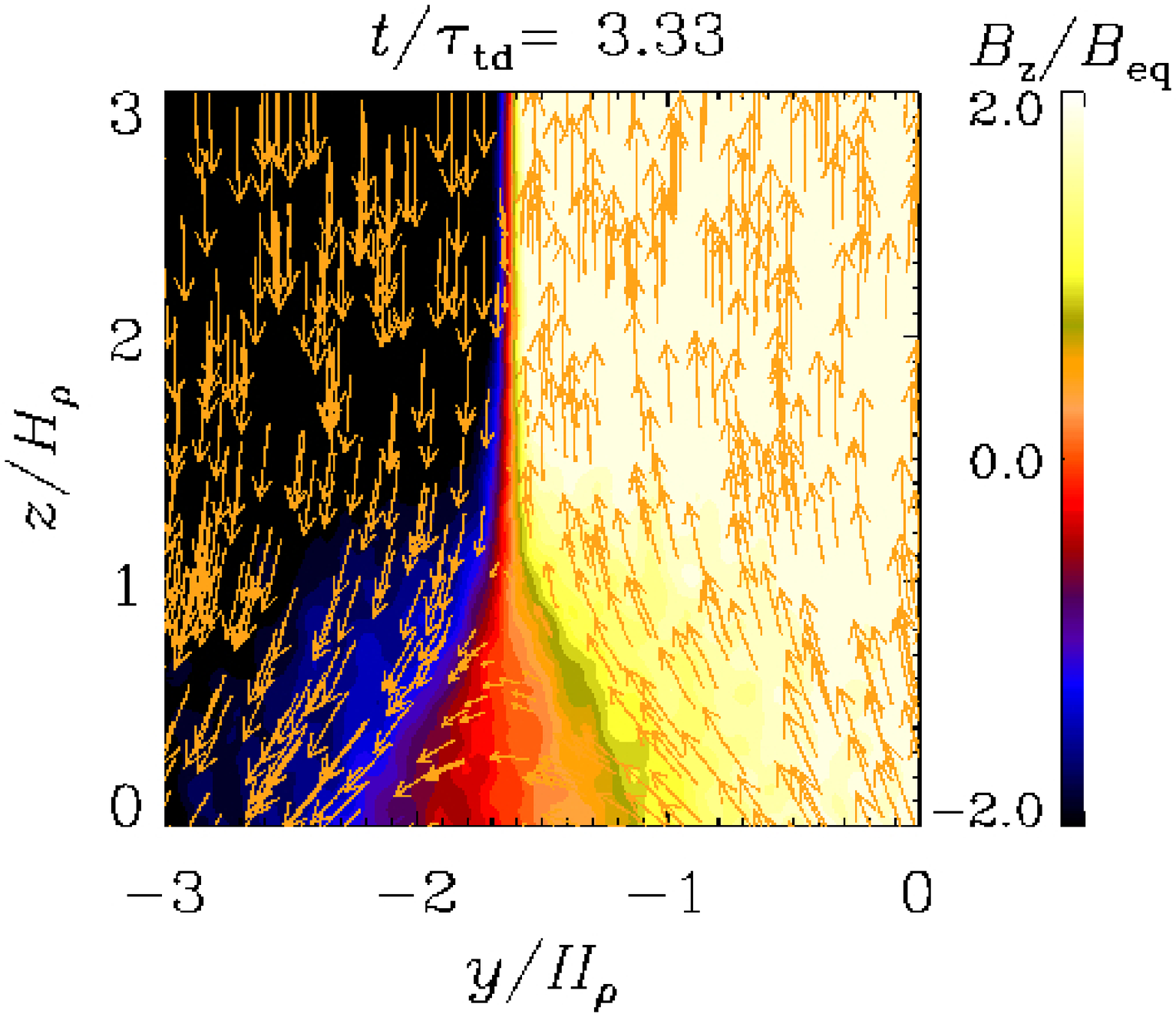}
\includegraphics[width=.32\columnwidth]{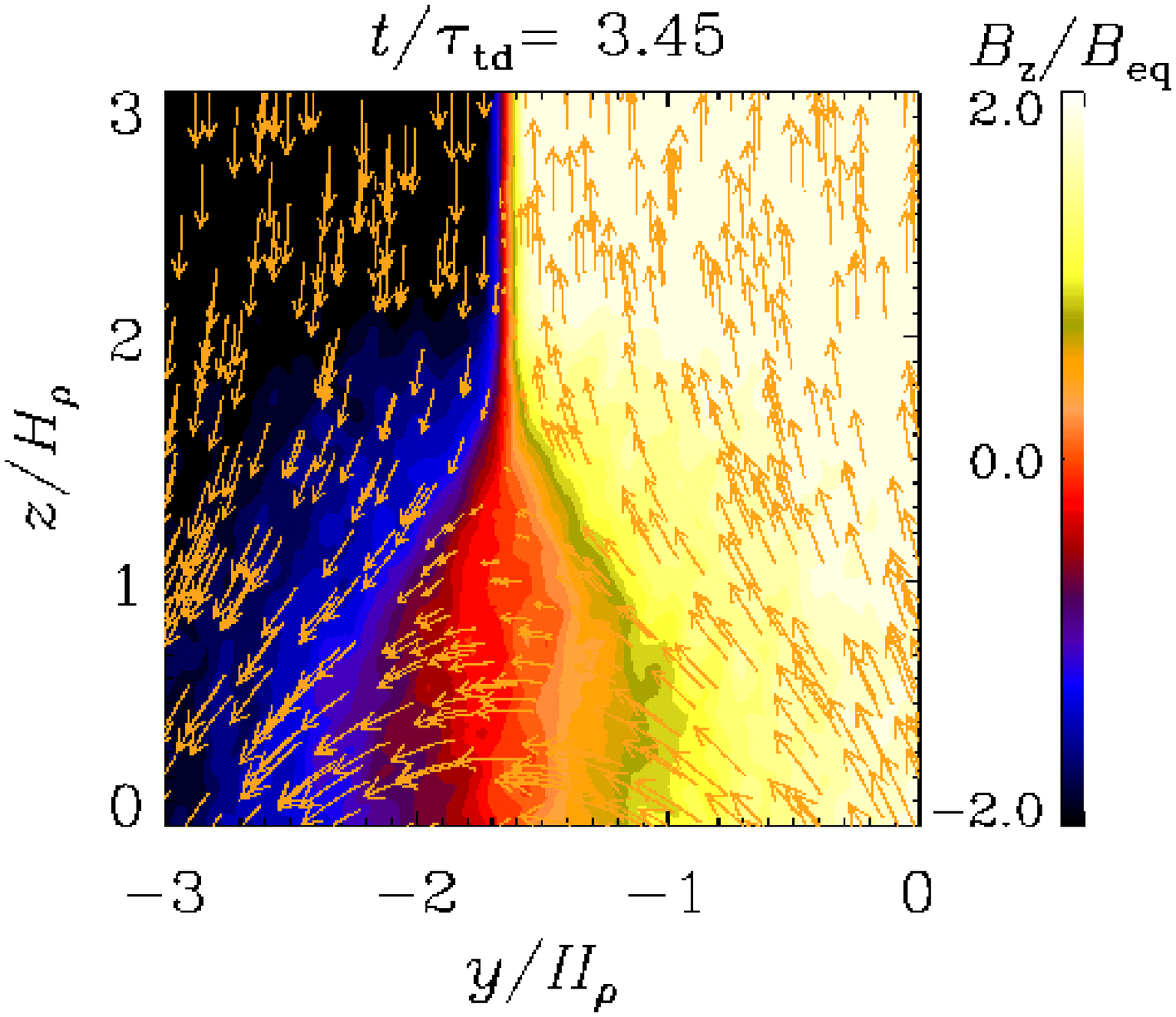}
\includegraphics[width=.32\columnwidth]{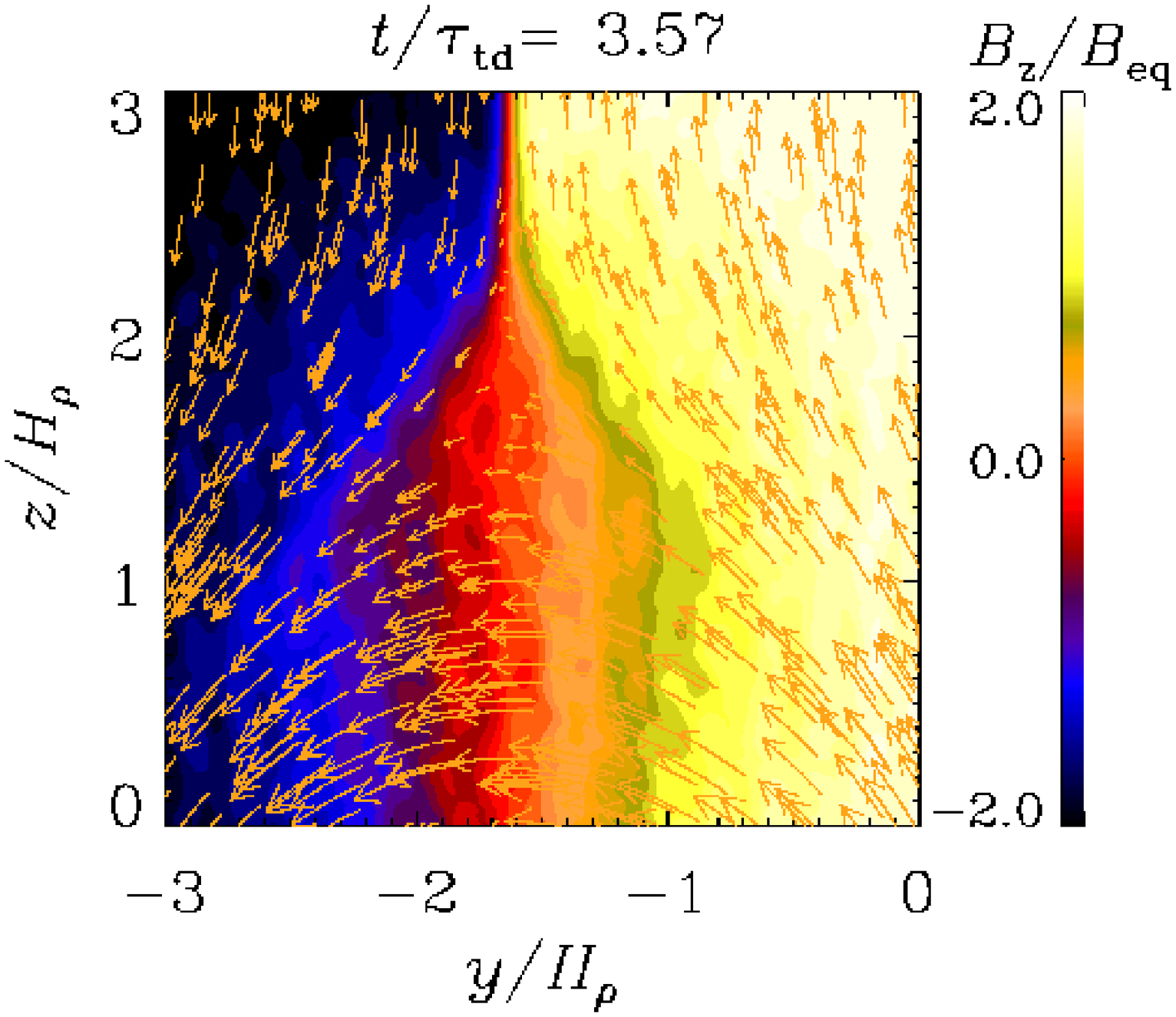}
\end{center}\caption[]{
Time evolution of $\mean{B_{z}}/\Beq$, together with $\mean{B_{y}}/\Beq$
and $\mean{B_{z}}/\Beq$ vectors for a case where the magnetic field is
generated by a helical dynamo.
Adapted from Ref.~\cite{JBMKR16}.
Reproduced from S.\ Jabbari et al.\ Turbulent Reconnection of Magnetic
Bipoles in Stratified Turbulence. MNRAS (2016) 459 (4): 4046-4056.
Published by Oxford University Press on behalf of the Royal Astronomical
Society online at:
http://mnras.oxfordjournals.org/content/459/4/4046.abstract?sid=5ba53a74-4653-4fbc-bc7d-58b9a4ffb816
}\label{pvarm}\end{figure*}

Although NEMPI does not appear to be excited when rotation is too strong,
simulations show that
magnetic flux concentrations are still being produced
when there is strong stratification and a dynamo is operating
preferentially in the deeper parts \cite{JBLKR14}.
In \Fig{pvarm} we show a case where a sub-equipartition strength magnetic
field is produced in the deeper parts and leads to a super-equipartition
strength magnetic fields in the surface layers.
The different polarities can then be driven together to form sharp
structures in the form of an inverted {\sf Y}-shaped pattern
in a vertical cross-section  \cite{JBMKR16,JBKR16}.
This is an example where the phenomenon of turbulent magnetic reconnection
has been seen to occur in a natural setting.
Similar behavior has also been seen in global spherical shell dynamos
\cite{JBKMR15}.
In simulations \cite{JBMKR16,JBKR16}, the generated magnetic field
reached super-equipartition levels so rapidly that it was not possible
to detect NEMPI during the growth of the magnetic field.
It should be noted that NEMPI cannot be excited for
super-equipartition magnetic fields.

To investigate the role of magnetic reconnection,
the flow around the sharp interface was zoomed in,
and the dynamics of the current sheet in this region
was studied \cite{JBMKR16}.
The reconnection rate was determined independently through
the inflow velocity in the vicinity of the current sheet
and via the electric field in the reconnection region.
For large Lundquist numbers ($\Lu>1000$), the reconnection rate was found
to be nearly independent of the value of $\Lu$ \cite{JBMKR16},
where $\Lu=v_{\rm A} L_c /\eta$ with $L_c$ being the length of the current sheet.
This agrees with earlier studies of turbulent reconnection \cite{LUS09},
which also showed independence of Ohmic resistivity \cite{LV99,LKT16},
as well as results of recent numerical simulations performed by other
groups in simpler settings \cite{LUS09,KLV09,HB10,LSS12,B13}.

\bigskip
\section{Turbulent convection}

In view of application to the Sun as a next step, it will be important to consider
more realistic modeling and include the effects of convectively-driven turbulence.
It is not obvious that NEMPI applies straightforwardly to this case.
It was shown already early on that---even in convection---the relevant
NEMPI parameter $\qp$ is indeed much larger than unity \cite{KBKMR12},
favoring the possibility of NEMPI.
Subsequent convection simulations with an imposed magnetic field
\cite{KBKKR16} yielded structures that are strongly reminiscent of
those found in realistic solar surface simulations in the presence of
full radiative transport \cite{SN12}.

The spontaneous formation of surface magnetic structures from a
large-scale $\alpha^2$ dynamo by strongly stratified thermal convection
in Cartesian geometry has recently also been studied by \cite{MS16}.
They found that large-scale magnetic structures are formed at the
surface only in cases with strong stratification.
The presence of rapid uniform rotation was an
argument in \cite{MS16} that NEMPI seems not to be responsible for
these structures.
On the other hand, the combined effect of rapid uniform rotation and
stratification can produce helicity and an $\alpha$ effect,
which causes a large-scale $\alpha^2$ dynamo.
A similar situation was encountered in connection with the dynamo
simulations of forced turbulence, where rotation also did not suppress
the formation of structures \cite{JBMKR16}.
Thus, the question of the origin of these structures remains unsettled.

The direct detection of negative effective magnetic pressure in
turbulent convection with dynamo-generated magnetic fields
is a difficult problem.
In many existing convection simulations,
unlike the case of forced turbulence, the scale separation
between the integral scale of the turbulence and the size of the domain
is not large enough for the excitation of NEMPI
and the formation of sharp magnetic structures.

\bigskip

\section{Cross helicity effect}

The role of NEMPI is not always evident, especially when the magnetic
field strongly exceeds the equipartition value.
However, all these simulations have in common that there is a vertical
magnetic field such that $\grav\cdot\meanBB\neq0$.
Interestingly, $\grav\cdot\meanBB$ is not only a pseudoscalar,
but it is odd in the magnetic field.
In MHD, there is an important invariant of the ideal equations,
namely the cross helicity $\bra{\uu\cdot\bm{B}}$ \cite{Wol58}.
It is often not important unless it was present in the initial conditions.
However, it is not too surprising that $\bra{\uu\cdot\bm{B}}$ is
generated whenever $\grav\cdot\meanBB$ is nonzero.
This was explored in Ref,~\cite{RKB11}, where it was found that
\EQ
\bra{\uu\cdot\bb}={\etat\over\cs^2}\,\grav\cdot\meanBB
=-{\etat\over H_\rho}\meanB_z.
\EN
If the turbulence intensity is nonuniform, there is yet another
contribution to $\bra{\uu\cdot\bb}$ that is proportional to
$\etat(\nab\ln\urms)\cdot\meanBB$, which
was first obtained in Ref.~\cite{KKMR03}.

\begin{figure}[t!]\begin{center}
\includegraphics[width=.8\columnwidth]{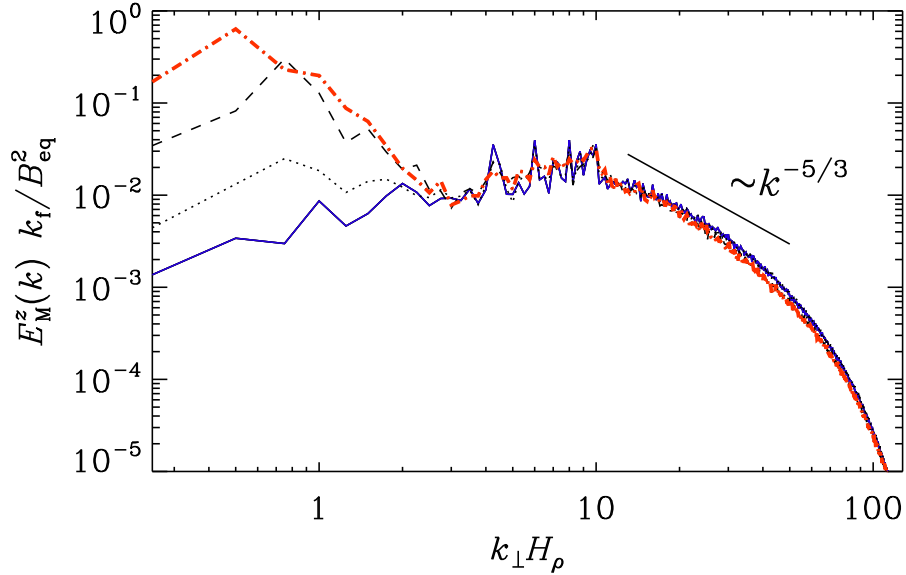}
\end{center}\caption[]{
Normalized spectra of $B_z$ from DNS at normalized times
$t\etatz/H_\rho^2\approx0.2$ (solid blue), 0.5 (dotted), 1 (dashed),
and 2.7 (dash-dotted red) with $\kf H_\rho=10$ and $k_1 H_\rho=0.25$.
Adapted from Ref.~\cite{BGJKR14}.
}\label{pBz_spec2_halfV1024x384k10VF_Bz02_g1_pm1}
\end{figure}

All the simulations that produce large-scale magnetic field structures
display what can be characterized as inverse cascade or
inverse transfer behavior of
magnetic field from the driving scale to large scales with a horizontal
wavenumber $k_\perp$ such that $k_\perp H_\rho\approx0.8$ or even less.
The role of the conservation property of $\bra{\uu\cdot\bm{B}}$
still needs to be explored, but it is clear that there is a
remarkable analogy between inverse transfer seen in
\Fig{pBz_spec2_halfV1024x384k10VF_Bz02_g1_pm1} and that of a
large-scale dynamo, where the conservation of magnetic helicity is
known to lead to an inverse cascade \cite{FPLM75,PFL76,Bra01}.

\section{Concluding remarks}

The purpose of this review has been to highlight the fact that the
presence of gravitational stratification introduces a qualitatively
new phenomenon in MHD turbulence, namely the formation of large-scale
magnetic structures
via excitation of NEMPI.
Turbulence causes a modification of the large-scale magnetic pressure,
so that the effective magnetic pressure becomes negative for large fluid and
magnetic Reynolds numbers, and this results in the excitation of NEMPI.
DNS demonstrate that the effective magnetic pressure can be
negative in any kind of turbulence, e.g., in non-stratified and stratified
isothermal turbulence, polytropic stably stratified turbulence,
turbulent convection, and in turbulence with an upper non-turbulent layer.
However, the actual instability is excited only in stratified turbulence
when the initial mean magnetic field is less than the equipartition field.
For very large fluid and magnetic Reynolds numbers,
NEMPI weakly dependent on the level of turbulence.
In some cases, where there is locally a vertical magnetic field,
NEMPI causes the formation of magnetic spots.

In astrophysics, there are many examples where MHD turbulence is
accompanied by strong density stratification.
A prediction from our studies reviewed in this paper would be
that such systems should exhibit magnetic spots.
We do not know whether there is a direct relation to sunspots, which are
generally hypothesized as being the result of deeply rooted thin
magnetic flux tubes.
Whether such isolated flux tubes really exist and how they are
formed is an open question.

We do know of magnetic flux tubes in MHD turbulence
\cite{NBJRRST92,BJNRST96}, that are analogous to vortex tubes in
hydrodynamic turbulence \cite{SJO90}, but these tend to scale with the
resistive scale \cite{BPS95}, so such tubes would become smaller as the
magnetic Reynolds number is increased.
Global simulations of convective spherical shell dynamos have been used
to visualize magnetic flux tubes \cite{NBBMT13,NBBMT14}.
Those may well be the type of tubes seen in earlier Cartesian simulations,
but they could also be local field enhancements resulting from the
large-scale dynamo.
It is hard to imagine that these flux structures alone can explain the
formation of sunspots, unless there was some kind of reamplification.
Clearly, as future simulations of global dynamos gain in resolution, they
would eventually display spots, just like the Sun and other stars do.
It will then be important to have possible frameworks in place for
understanding such spots.
We hope that the present review has provided some relevant inspiration
beyond the standard paradigm.

Further steps toward more realistic modeling
of the formation of magnetic spots and bipolar regions
include replacing forced turbulence by self-consistently
driven convective motions that are influenced by the radiative
cooling at the surface together with partial ionization.
Including more realistic physical processes at the solar surface might
also help to reproduce the surrounding spot structures.

\bigskip
\noindent
{\bf Acknowledgments}
\bigskip

IR and NK thank NORDITA for hospitality and support during their visits.
This work has been supported in parts by the Swedish Research Council
grant No.\ 621-2011-5076 and the Research Council of Norway under the
FRINATEK grant No.\ 231444.

\bigskip
\noindent
{\bf References}
\bigskip


\vfill\bigskip\noindent\tiny\begin{verbatim}
$Header: /var/cvs/brandenb/tex/nathan_igor/NEMPI/paper.tex,v 1.98 2016/12/24 14:34:04 brandenb Exp $
\end{verbatim}
\end{document}